# Bacteria solve the problem of crowding by moving slowly

O. J. Meacock[1,2]†, A. Doostmohammadi[3]†, K. R. Foster[1]*, J. M. Yeomans[3]*, W. M. Durham[1,2]*

[1] Department of Zoology, South Parks Road, University of Oxford, Oxford, OX1 3PS, United Kingdom

[2] Department of Physics and Astronomy, University of Sheffield, Hounsfield Road, Sheffield, S3 7RH, United Kingdom

[3] Rudolf Peierls Centre for Theoretical Physics, Clarendon Laboratory, Parks Road, University of Oxford, Oxford, OX1 3PU, United Kingdom

* e-mail: kevin.foster@zoo.ox.ac.uk, julia.yeomans@physics.ox.ac.uk, or w.m.durham@sheffield.ac.uk

†These authors contributed equally.

**In systems as diverse as migrating mammals to road traffic, crowding acts to inhibit efficient collective movement**[1–4]**. Bacteria, however, are observed to move in very dense groups containing billions of individuals without causing the gridlock common to other systems**[5,6]**. Here we combine experiments, cell tracking and individual-based modelling to study the pathogen** *Pseudomonas aeruginosa* **as it collectively migrates across surfaces using grappling-hook like pili**[6–8]**. We show that the fast moving cells of a hyperpilated mutant are overtaken and outcompeted by the slower moving wild-type at high cell densities. Using theory developed to study liquid crystals**[9–14]**, we demonstrate that this effect is mediated by the physics of topological defects, points where cells with different orientations meet one another. Our analyses reveal that when comet-like defects collide with one another, the fast-moving mutant cells rotate vertically and become trapped. By moving more slowly, wild-type cells avoid this trapping mechanism, allowing them to collectively migrate faster. Our work suggests that the physics of liquid crystals has played a pivotal role in the evolution of collective bacterial motility by exerting a strong selection for cells that exercise restraint in their movement.**

Patterns of collective motility are shaped by the tension between the interests of individuals and those of the group[15] – while selfish behaviours can potentially benefit individuals, they often have catastrophic consequences if adopted by a larger proportion of the population[16]. Familiar examples from our own society include automobile traffic jams and crushes in groups of panicked pedestrians[1,2]. Although these principles seem especially relevant to bacteria, which move across surfaces in very densely packed collectives[5,6], how the interests of individual bacteria and those of the group are balanced remain unexplored.

Bacteria generate collective motility on surfaces through a variety of mechanisms[6,17]. One of the best studied is twitching motility, where cells pull themselves across surfaces using grappling hook-like appendages called Type-IV pili[6–8]. Bacteria can also move using flagella, but cells swimming through liquid with flagella must be concentrated by orders of magnitude for collective behaviours to emerge, and these are typically disrupted within minutes by oxygen depletion[18,19]. Here we study the relationship between individual and collective twitching motility within large groups of the opportunistic pathogen *P. aeruginosa*. We focus on movement in subsurface colonies (Fig. 1a) where we can use high-resolution, time-lapse imaging to track individual cells as they use pili to collectively move along a glass surface (Extended Data Fig. 1). We also find that the key features of this assay are reproduced in classical bacterial colonies grown on the surface of agar (Fig. 1a-d, Extended Data Fig. 2, Supplementary Movie 1).

Recent microfluidic experiments have shown that deleting the *pilH* gene, which encodes a response regulator in the two component system that regulates twitching motility[20], causes cells to move faster than wild-type (WT) cells[21]. To confirm that this increase also occurs in subsurface colonies, we developed a cell tracking algorithm to record the movement of thousands of individual cells within a single field of view (Methods, Supplementary Movie 2).

This revealed that individual Δ*pilH* cells move approximately twice as fast as WT cells in both densely packed collectives (Fig. 1g) and at lower cell densities (Extended Data Fig. 3b). However, we were surprised to find that individual cell speed did not translate to how quickly colonies spread across the surface. While the faster motility of Δ*pilH* cells allowed them to initially spread outwards faster than WT cells, their expansion rate plateaued after four hours. In contrast, the expansion rate of WT colonies steadily increased over a period of approximately six hours, eventually reaching a value four times greater than that of Δ*pilH* colonies (Fig. 1e, f, Supplementary Movie 3). An increase in individual motility, therefore, did not translate successfully to an increase in collective motility.

Previous work has shown that strains which dominate the nutrient-rich edge of colonies can obtain a substantial fitness advantage[22]. To investigate how this effect might impact fitness of Δ*pilH* cells, we directly competed Δ*pilH* and WT cells by combining them in surficial colonies. This revealed that Δ*pilH* cells remained trapped in the nutrient poor interior of the colony, while WT cells migrated outwards (Fig. 2a), allowing the latter to undergo approximately three more cell divisions over a 48-hour period (Fig. 2b). We next analysed the dynamics of this competition in greater detail using subsurface colonies and automated image analysis to simultaneously measure colony expansion rate, cell packing fraction, and genotypic composition (Fig. 2c-g, Extended Data Fig. 4, Supplementary Movie 4). Consistent with their initial rapid expansion in monoculture colonies, Δ*pilH* cells initially outnumbered WT cells at the expanding edge of the colony (hereafter "the front"). However, once the front transitioned from loosely packed groups of cells to a confluent monolayer (Fig. 2e, g), the colony expansion rate rapidly increased from 0.75 µm min$^{-1}$ to 5.0 µm min$^{-1}$ (Fig. 2d) while the proportion of Δ*pilH* cells at the front fell sharply, dropping from 88% at 200 min to 11% at 400 min (Fig. 2f).

What could be responsible for this rapid decline of Δ*pilH* despite its faster individual movement? Hyperpilation can increase both doubling times[23] and cell-cell adhesion[23,24], but our analyses show that neither effect is sufficient to explain the sudden decline of the fraction of Δ*pilH* cells at the colony front (Supplementary Notes, Extended Data Figs. 5-7). Instead, we hypothesized the precipitous decline of fast moving Δ*pilH* cells at the front stemmed from an inability to move collectively. To understand how the collective behaviours of the two strains differ, we turned to tools originally developed to study liquid crystals. Neighbouring cells in colonies are closely aligned with one another, and this "nematic" ordering produces collective movement on length scales substantially larger than that of a single cell[25,26]. Moreover, we observe that topological defects, an emergent feature of nematic systems, move about within the monolayer of cells (Supplementary Movie 5, Methods). The two types of defects – denoted here as "comets" and "trefoils" (Fig. 3a-c) – are generated and annihilated in pairs (Extended Data Fig. 8a, b), as predicted by theory[9,10].

To investigate the physical properties of defects, we developed automated tools to combine single-cell tracking data collected across hundreds of defects. This revealed that the movements of both WT cells (Fig. 3d) and Δ*pilH* cells (Extended Data Fig. 8c) around comets and trefoils closely match predictions from both an individual-based model of self-propelled rods (SPR)[18] and a continuum model of active nematics[11], but that the Δ*pilH* flowfields are larger in scale than those of the WT. Theory predicts that comets migrate along their axis at a speed proportional to the "activity" of the nematic, a measure of the force exerted by each of the individuals that make up the system. In contrast, trefoils are predicted to move diffusively[12]. Consistent with this, we observed that the root mean squared displacement (RMSD) of comets was larger in Δ*pilH* monolayers than in WT monolayers, while the RMSD of trefoils was similar in both genotypes (Fig. 3e, f). These observations indicate that both WT and Δ*pilH* monolayers behave as an active nematic, with the latter possessing greater activity.

Comets repel one another in a nematic confined to two dimensions[13]. However, previous theoretical predictions[10,14] and experiments with liquid crystals[14] suggested that when the nematic is allowed to reorient out of the 2D plane, two comets at a sufficiently small separation could merge together, causing the cells within the comets to "escape into the third dimension" by standing up vertically. We hypothesized that this process allows the fast-moving comets in our experiments to merge together, causing the higher activity Δ*pilH* cells to become trapped in place. To test this possibility, we first extended our SPR model to three dimensions to allow rods to reorient out of the plane and simulated collisions between comets (Methods). Stable structures of upright rods ("rosettes") formed once the rods' propulsive force, $F$, increased beyond a critical threshold, $F_v$, confirming our intuition (Fig. 4a, Extended Data Fig. 10, Supplementary Movies 6, 7). Δ*pilH* cells are also slightly longer than WT cells (Extended Data Fig. 6a), but this acts to suppress the nucleation of rosettes rather than promote it (Extended Data Fig. 6b-e, Supplementary Notes). To test if the increased force generated by Δ*pilH* cells is alone sufficient to preferentially trap them in rosettes, we next used our 3D SPR model to simulate the interaction of two different genotypes that each exert a different propulsive force (Fig. 4b, c). This showed that higher-force mutants, on average, move more slowly than the WT cells because a larger fraction of the higher-force mutants become stuck in rosettes where their movement is arrested.

We tested these predictions by inoculating subsurface colonies composed of a mixture of both Δ*pilH* and WT cells. Δ*pilH* cells spontaneously formed aggregations in these colonies, whereas WT cells did not (Supplementary Movie 8). Moreover, we were able to quantify the movement of topological defects at precisely the time when the fraction of Δ*pilH* cells in the front sharply decreases (Fig. 2f). We observed comets approaching each other (Fig. 4d) before the monolayer of cells buckled to generate a rosette (Fig. 4e, Extended Data Fig. 10, Supplementary Movie 9), as predicted by our SPR model (Fig. 4a). Even though this colony was initiated with an equal fraction of WT and Δ*pilH* cells, confocal imaging revealed that the core of this rosette was nearly entirely composed of vertically oriented Δ*pilH* cells (Fig. 4f). Once initiated, rosettes in both experiments and simulations grew larger, similar to the "inverse domino" cascade of cell verticalization observed at the centre of non-motile bacterial colonies[27–29].

Our results indicate that the physical processes that control the movement of crowds of bacteria exert a fundamental speed limit on cell motility. Cells that exceed this critical threshold form high-velocity comets that are unstable to verticalization upon collision, and this causes fast moving cells to become trapped within the interior of colonies where nutrients are scarce[30]. Thus, bacteria collectively benefit by moving more slowly as individuals. Collective benefits are often insufficient to prevent the problems caused by crowding in other systems, including human society, because being the first to move more rapidly in a crowd can provide an individual with substantial benefits[1–3,16]. However, we have shown bacterial collectives have solved this wide-spread dilemma: fast moving cells briefly move ahead of the pack, before crashing into one another and falling foul of their own strategy. In this way, natural selection acts to prioritize efficient collective migration by favouring individual cells that exercise restraint in their movement.

1. Helbing, D., Farkas, I. & Vicsek, T. Simulating dynamical features of escape panic. *Nature* **407**, 487–490 (2000).
2. Helbing, D. Traffic and related self-driven many-particle systems. *Rev. Mod. Phys.* **73**, 1067–1141 (2001).
3. Saloma, C., Perez, G. J., Tapang, G., Lim, M. & Palmes-Saloma, C. Self-organized queuing and scale-free behavior in real escape panic. *PNAS* **100**, 11947–52 (2003).
4. Cates, M. E. & Tailleur, J. Motility-Induced Phase Separation. *Annu. Rev. Condens. Matter Phys.* **6**, 219–244 (2015).

5. Zhang, H. P., Be'er, A., Florin, E.-L. & Swinney, H. L. Collective motion and density fluctuations in bacterial colonies. *PNAS* **107**, 13626–30 (2010).

6. Gloag, E. S. *et al.* Self-organization of bacterial biofilms is facilitated by extracellular DNA. *PNAS* **110**, 11541–6 (2013).

7. Talà, L., Fineberg, A., Kukura, P. & Persat, A. Pseudomonas aeruginosa orchestrates twitching motility by sequential control of type IV pili movements. *Nat. Microbiol.* (2019). doi:10.1038/s41564-019-0378-9

8. Skerker, J. M. & Berg, H. C. Direct observation of extension and retraction of type IV pili. *PNAS* **98**, 6901–4 (2001).

9. Doostmohammadi, A., Ignés-Mullol, J., Yeomans, J. M. & Sagués, F. Active nematics. *Nat. Commun.* **9**, 3246 (2018).

10. Mermin, N. D. The topological theory of defects in ordered media. *Rev. Mod. Phys.* **51**, 591–648 (1979).

11. Giomi, L., Bowick, M. J., Mishra, P., Sknepnek, R. & Cristina Marchetti, M. Defect dynamics in active nematics. *Philos. Trans. A.* **372**, 20130365 (2014).

12. Giomi, L., Bowick, M. J., Ma, X. & Marchetti, M. C. Defect Annihilation and Proliferation in Active Nematics. *PRL* **110**, 209901 (2013).

13. Gennes, P. G. de. & Prost, J. *The physics of liquid crystals*. (Clarendon Press, 1993).

14. Meyer, R. B. On the existence of even indexed disclinations in nematic liquid crystals. *Philos. Mag.* **27**, 405–424 (1973).

15. Hamilton, W. D. Geometry for the selfish herd. *J. Theor. Biol.* **31**, 295–311 (1971).

16. Gershenson, C. & Helbing, D. When slower is faster. *Complexity* **21**, 9–15 (2015).

17. Jarrell, K. F. & McBride, M. J. The surprisingly diverse ways that prokaryotes move. *Nat. Rev. Microbiol.* **6**, 466–476 (2008).

18. Wensink, H. H. *et al.* Meso-scale turbulence in living fluids. *PNAS* **109**, 14308–13

(2012).

19. Dunkel, J. *et al.* Fluid Dynamics of Bacterial Turbulence. *PRL* **110**, 228102 (2013).

20. Bertrand, J. J., West, J. T. & Engel, J. N. Genetic Analysis of the Regulation of Type IV Pilus Function by the Chp Chemosensory System of Pseudomonas aeruginosa. *J. Bacteriol.* **192**, 994–1010 (2009).

21. Oliveira, N. M., Foster, K. R. & Durham, W. M. Single-cell twitching chemotaxis in developing biofilms. *PNAS* **113**, 6532–6537 (2016).

22. Kim, W., Racimo, F., Schluter, J., Levy, S. B. & Foster, K. R. Importance of positioning for microbial evolution. *PNAS* **111**, E1639-47 (2014).

23. Oldewurtel, E. R., Kouzel, N., Dewenter, L., Henseler, K. & Maier, B. Differential interaction forces govern bacterial sorting in early biofilms. *Elife* **4**, e10811 (2015).

24. Anyan, M. E. *et al.* Type IV pili interactions promote intercellular association and moderate swarming of Pseudomonas aeruginosa. *PNAS* **111**, 18013–8 (2014).

25. Dell'Arciprete, D. *et al.* A growing bacterial colony in two dimensions as an active nematic. *Nat. Commun.* **9**, 4190 (2018).

26. Doostmohammadi, A., Thampi, S. P. & Yeomans, J. M. Defect-Mediated Morphologies in Growing Cell Colonies. *PRL* **117**, 048102 (2016).

27. Beroz, F. *et al.* Verticalization of bacterial biofilms. *Nat. Phys.* **14**, 954–960 (2018).

28. Drescher, K. *et al.* Architectural transitions in Vibrio cholerae biofilms at single-cell resolution. *PNAS* **113**, E2066-72 (2016).

29. Grant, M. A. A., Wacław, B., Allen, R. J. & Cicuta, P. The role of mechanical forces in the planar-to-bulk transition in growing Escherichia coli microcolonies. *J. R. Soc. Interface* **11**, (2014).

30. Pirt, S. J. A Kinetic Study of the Mode of Growth of Surface Colonies of Bacteria and Fungi. *Microbiology* **47**, 181–197 (1967).

**Acknowledgements** We thank S. Booth, N. Clarke, C. Durham, A. Fenton, E. Granato, J. Guasto, J. Hobbs, T. Meiller-Legrand, W. Smith, A. Morozov and J. Wheeler for providing comments on a previous version of this manuscript, R. Allen and R. Hawkins for helpful discussions, W. Smith for assistance with the SPR model, J. Engel for sharing bacterial strains, and M. Hopkins for assistance with preliminary experiments. O.J.M. was supported by an EPSRC studentship through the Life Sciences Interface Centre for Doctoral Training (EP/F500394/1), A.D. was supported by the Royal Commission for Exhibition of 1851 Research Fellowship, K.R.F. was supported by European Research Council Grant 787932 and a Wellcome Trust Investigator award, and W.M.D was supported by a startup grant from the University of Sheffield's Imagine: Imaging Life initiative, an EPSRC Pump Priming Award (EP/M027430/1) and a BBSRC New Investigator Grant (BB/R018383/1).

**Author contributions** O.J.M. performed experiments, implemented the SPR model, analysed data, and prepared figures. A.D. and J.M.Y. proposed the mechanism of rosette formation. O.J.M, A.D., K.R.F, J.M.Y., and W.M.D. all contributed to the design of experiments and models, as well as to the interpretation of results. O.J.M, K.R.F, and W.M.D. wrote the paper with input from A.D. and J.M.Y. This collaborative effort was led by W.M.D.

**Author information** The authors declare no competing interests. Correspondence and requests for materials should be addressed to <u>W.M.D, K.R.F or J.M.Y.</u>

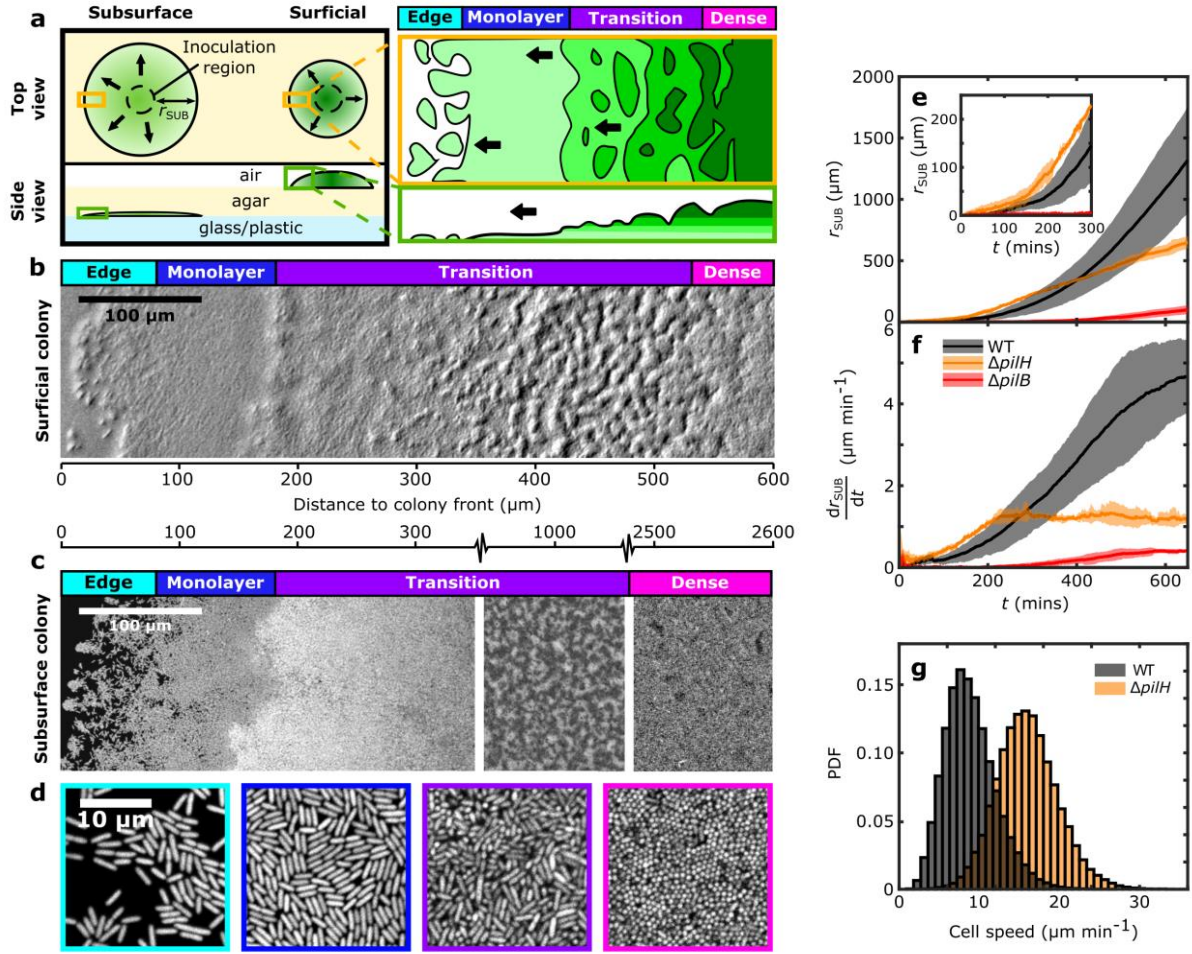

**Fig. 1 | Pili-based motility drives the spread of *P. aeruginosa* colonies, but cells that individually move faster spread more slowly**. **a**, Both subsurface colonies, which grow beneath a layer of agar, and surficial colonies, which grow on top of agar, consist of four distinct regions. **b**, **c**, **d**, The "edge" contains small groups of loosely packed cells, the "monolayer" is composed of tightly packed cells lying flat against surface, the "transitional" region is a mixture of horizontally and vertically oriented cells, and in the "dense" region almost all cells are vertical. Panel **d** shows magnified views of each subsurface colony region. **e**, **f**, Measurements of the subsurface colony radius, $r_{SUB}$, and expansion rate, $dr_{SUB}/dt$ for monocultures of WT (black), Δ*pilH* (orange) and non-piliated Δ*pilB* (red) cells. Inset in **e** shows a magnified view of the first 300 mins. Shaded regions in **e**, **f** show the standard deviation about the mean for three separate experiments. **g**, The probability density function (PDF) of the speed of individual WT and Δ*pilH* cells within the monolayer region.

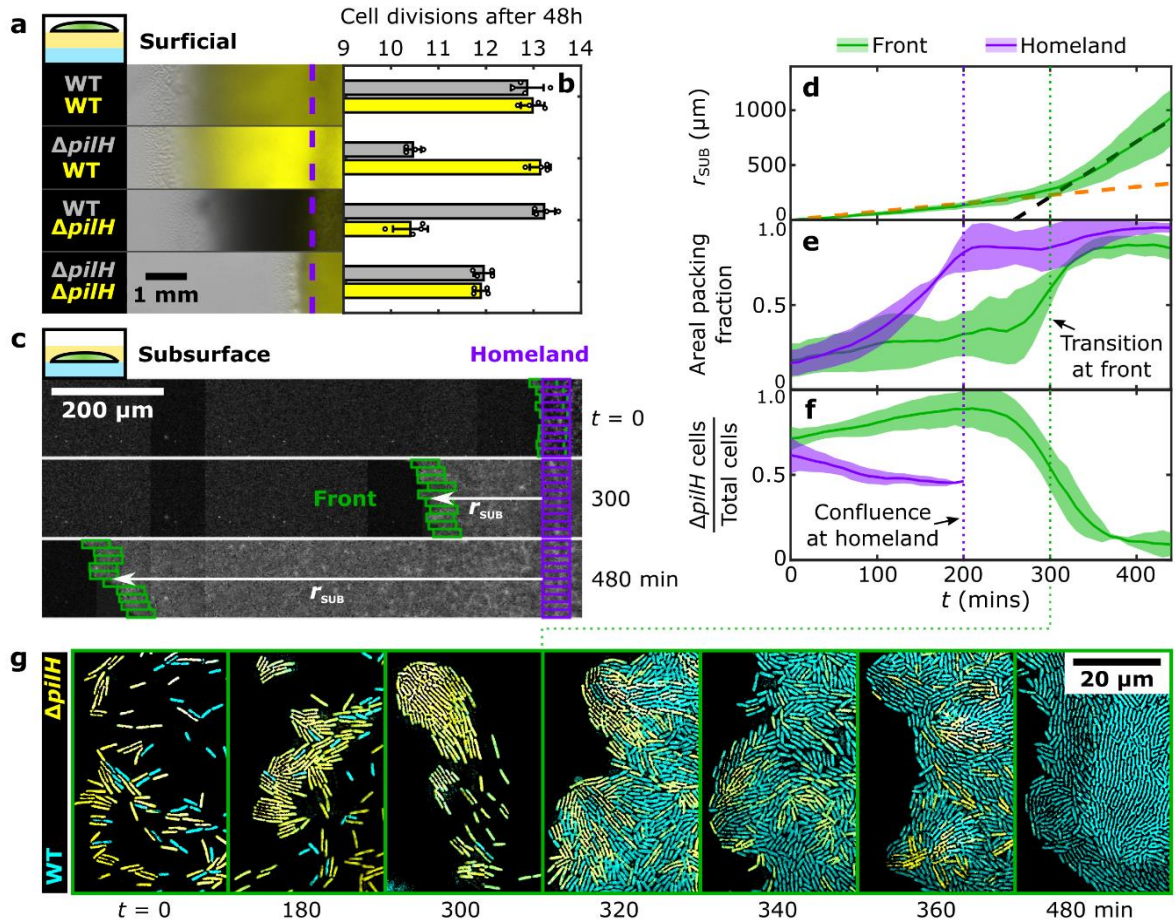

**Fig. 2 | WT cells outcompete ΔpilH cells when mixed together in the same colony. a**, Surficial colonies inoculated with equal proportions of YFP and CFP labelled cells after 48 hours. Dashed purple lines indicate the boundary of the "homeland", the region where cells are initially inoculated onto the agar. Due to imaging constraints (Methods), we show only the YFP and brightfield channels. Both strains appear in the latter, so regions with more CFP-labelled cells appear darker. **b**, Number of cell divisions occurring over 48 hours of incubation within colonies shown in **a**. Grey and yellow bars denote populations labelled with CFP and YFP, respectively. Error bars show the standard deviation of 4 replicates. **c**, Subsurface colonies were initialized with equal proportions of YFP labelled Δ*pilH* and CFP labelled WT cells. We analysed dynamics within both the homeland (stationary, purple boxes) and the "front" of the colony, which follows the edge of the colony has it expands (green boxes, Supplementary Movie 4). **d**, Measurement of the distance between the front and homeland regions, $r_{SUB}$, reveals a seven-fold increase in colony expansion rate at $t = 300$ mins (dashed lines, fitted with piecewise linear regression). The areal packing fraction (the fraction of the surface covered by cells) (**e**), and the relative frequency of Δ*pilH* cells (**f**) at the front also show sharp transitions at $t = 300$ mins. Shading in **d-f** shows the standard deviation from three experiments. Upon reaching confluence at $t = 200$ mins, cells in the homeland became too tightly packed to resolve their identity. **g**, Magnified images of the leading edge of a subsurface colony initiated with equal fractions of WT (cyan) and Δ*pilH* (yellow) cells. These images were processed as described in the Methods.

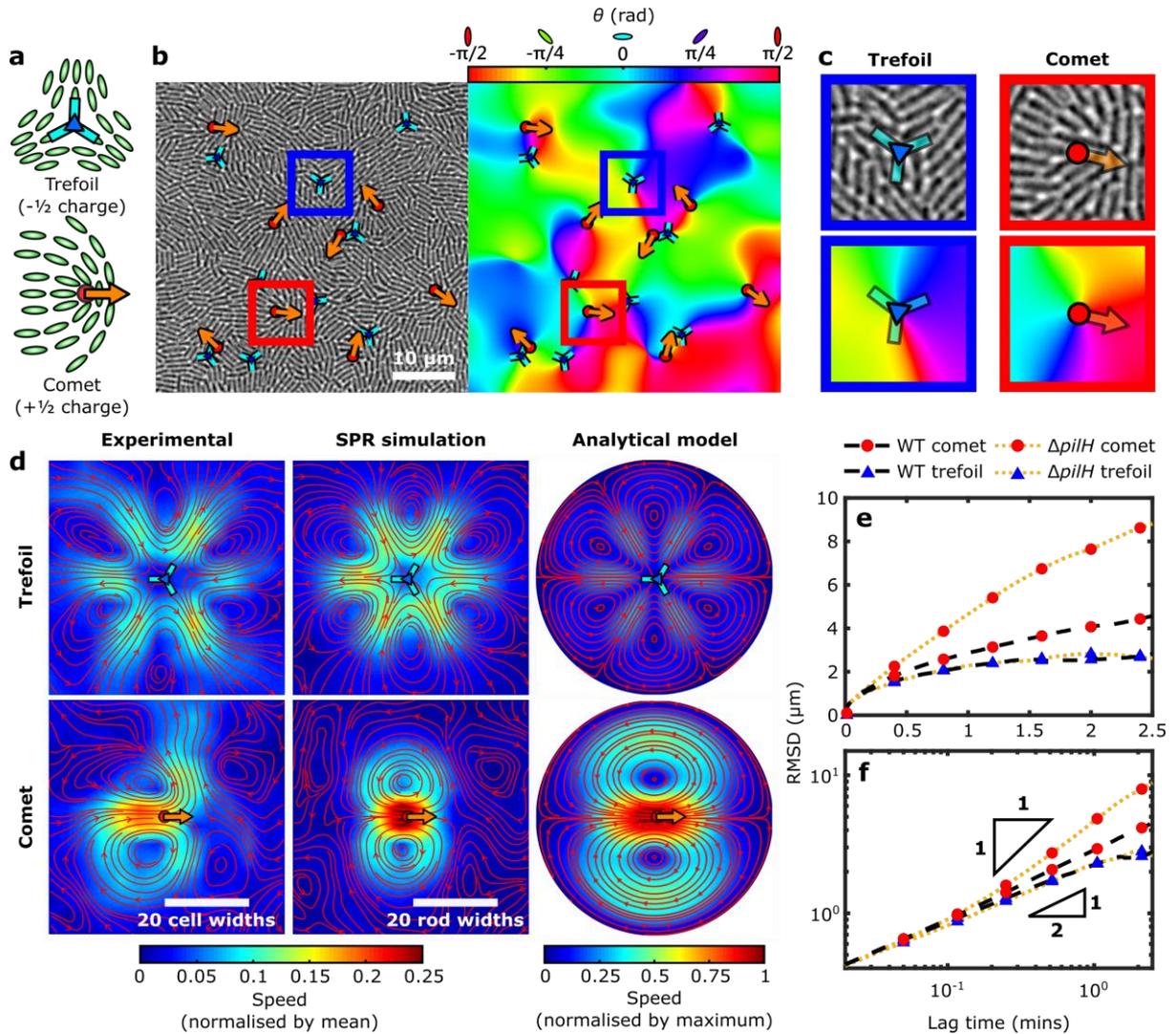

**Fig. 3 | Subsurface colonies exhibit patterns of collective motility consistent with a nematic system driven out of equilibrium. a**, Two types of topological defects occur in 2D nematic systems: trefoils, which exhibit three-fold rotational symmetry and have a topological charge of -1/2, and comets, which migrate along their single axis of symmetry and have a charge of +1/2 (see text). **b**, The locations and orientations of defects in the monolayer of a WT subsurface colony (left, Supplementary Movie 5) were obtained by quantifying the local cell orientation (right). **c**, Magnified views of the red and blue boxes in **b** illustrate how defects occur at singularities in cell orientation. **d**, Simultaneous tracking of both defects and individual cells within a monolayer (Supplementary Movies 2, 5) allows the mean cell flow around defects to be resolved. The structure of the flow closely resembles that predicted from a self-propelled rod (SPR) simulation and an analytical model (Methods). Red lines show streamlines and the background colours indicate flow speed. **e**, **f**, The root mean squared displacement (RMSD) measures how far defects move over a given lag time, plotted here on both a linear (**e**) and logarithmic scale (**f**). Triangles in **f** show the slopes predicted for ballistic (1:1) and diffusive (1:2) movement.

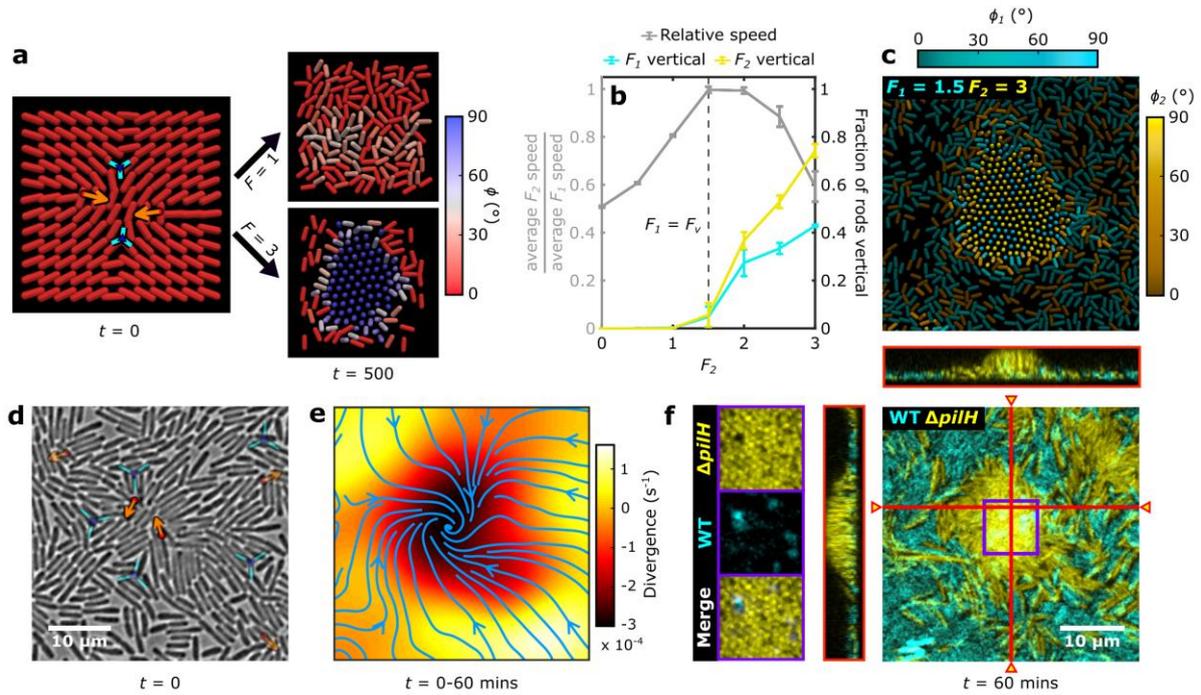

**Fig. 4 | Δ*pilH* cells are preferentially trapped by rosettes, preventing their outward migration. a**, A 3D SPR simulation of two comets colliding with one another shows that rods propelled by smaller force ($F=1$) remain flat against the surface after an initial transient, while those propelled by a larger force ($F=3$) form a stable, vertically oriented rosette (Supplementary Movie 6). Colour denotes the angle of rods relative to the surface, $\phi$. **b**, SPR simulations initialized with randomly oriented rods, half of which are propelled with a fixed force of $F_1=F_v=1.5$ (cyan, "WT") and the other half by a variable force $F_2$ (yellow, "mutant"). The left axis shows the mean speed of the $F_2$ population at steady state normalized by that of the $F_1$ population. The right axis indicates the fraction of each population that is vertically oriented (defined as $\phi > 85°$). Error bars indicate standard deviation of three different simulations, each with a different random initial configuration. **c**, A rosette from a $F_1=1.5$, $F_2 = 3$ simulation at steady state, where $\phi_1$ and $\phi_2$ denote the orientation of the two respective populations. **d**, Two comets approaching one another in a colony initiated with an equal number of WT and Δ*pilH* cells (Supplementary Movie 9). **e**, Measurements of cell velocity within same region as **d** during rosette development. Blue arrows and colormap respectively show streamlines and divergence of the time-averaged flow field (Methods). Regions of negative divergence indicate zones of cell accumulation. **f**, Three-dimensional confocal image of the rosette that formed within the same region shown in **d** and **e**, taken 60 mins after **d** (Methods). Vertical slices through the rosette are shown to the top and right, at locations indicated by triangles. (**f, inset**) A magnified view of the rosette (purple box in **f**) shows is it mostly composed of Δ*pilH* cells in a vertical orientation. Main panel shows maximal *z*-projection of both YFP and CFP channels, while insets show individual *z* slices.

# Methods

## Experimental Protocols

**Bacterial strains and fluorescent labelling**

The clean deletion mutants and the corresponding WT strains[20,31] used in this study were labelled with either CFP or YFP using a Gm$^r$ mini-Tn7 vector[32], using a three-strain mating protocol. Briefly, *P. aeruginosa* colonies were grown overnight at 42°C on an LB (Lennox, 20 g/l, Fisher Scientific) plate containing 1.5% (w/v) agar (Difco brand, BD). These were then mixed with both the mini-Tn7 donor *E. coli* strain and a SM10 λpir *E. coli* helper strain on the surface of a fresh LB agar plate. The resulting mixed three strain plate was incubated overnight at 30°C. Cells were then resuspended in liquid LB and transformants selected on LB agar plates containing both gentamicin (30 mg/l) and kanamycin (25 mg/l). The resulting CFP-and YFP-labelled strains were then directly competed with unlabelled strains to confirm that the impact of the labelling process on growth rate and motility was negligible (Extended Data Fig. 7).

**Bacterial cell culture**

We streaked -80°C freezer stocks onto LB agar plates and incubated them overnight at 37°C. Single colonies were picked and grown overnight in liquid LB at 37°C under continuous shaking. The following day, overnight cultures were diluted 30-fold in fresh LB broth and returned to the 37°C shaking incubator for two hours to obtain cells in exponential phase. Immediately before being used in colony experiments, the optical density at 600 nm ($OD_{600}$) was adjusted to 0.05 using fresh LB. For co-culture experiments, the optical densities of cultures of each individual strain were adjusted to $OD_{600} = 0.05$. These were then mixed in a single tube to ensure both strains were present in equal proportion. All colony-based assays were conducted at room temperature.

**Microscopic imaging**

Brightfield imaging of single genotype subsurface colonies was performed using a Zeiss Axio Observer inverted microscope ("Zeiss inverted") outfitted with a Zeiss MRm camera, Definite Focus system, Zeiss Zen software and either a 20X Plan Apochromat air objective or 63X Plan Apochromat oil-immersion objective. For co-culture experiments, both brightfield and epifluorescence imaging were performed using the same system, using a Zeiss HXP 120 illuminator for excitation. Confocal imaging was performed using a Zeiss LSM 700 laser scanning attachment, using a 63X Plan Apochromat oil-immersion objective for subsurface colonies and a 50X EC Epiplan Neofluar air objective for surficial colonies.

For other experiments, we used a Nikon Ti-E inverted microscope ("Nikon inverted") outfitted with a Perfect Focus System, a Plan Apochromat 100X brightfield objective, a Hamamatsu Flash 4.0 v2 camera and NIS-Elements software. We further increased the magnification using the microscope body's 1.5X zoom feature, yielding a total magnification of 150X.

Images of surficial colonies were taken with a Zeiss Axio Zoom.V16 zoom microscope ("Zeiss zoom"), outfitted with a Zeiss MRm camera, Zeiss Zen software and either a PlanApo Z 0.5X objective or a PlanNeoFluar Z 2.3X objective. Excitation of fluorophores was achieved using a Zeiss HXP 200 C illuminator.

**Surficial colony competition assay**

We initialized surficial colonies (Fig. 2a, b) using mixtures of the following strain pairs: Δ*pilH*–YFP/Δ*pilH*–CFP, Δ*pilH*–YFP/WT-CFP, WT-YFP/Δ*pilH*–CFP and WT-YFP/WT-CFP. We inoculated 10 µl of each mixture onto a freshly poured 1.5% (w/v) LB agar plate and sealed the plate lids with Parafilm (Bemis) to prevent evaporation. These were incubated at room temperature for 48 h.

We calculated the average number of cell divisions in each strain over the 48 hours of competition using the expression

$$\text{number of cell divisions} = \log_2 \left( \frac{\text{number of cells of strain inoculated onto surface}}{\text{number of cells of strain in colony after 48 h}} \right). \quad (1)$$

We measured the number of cells of each strain inoculated onto the surface by diluting the mixed liquid cultures used for inoculation and then spreading them onto LB plates. The resulting colonies were counted manually the after incubation overnight using the Zeiss zoom microscope, which allowed us to distinguish YFP and CFP expressing colony forming units (CFUs).

A similar technique was used to estimate the number of cells of each strain after 48 h of incubation. Whole surficial colonies were scraped, resuspended in fresh media, and vortexed. The resulting suspensions were then diluted, spread onto LB plates, and incubated overnight. We again used the Zeiss zoom microscope to manually enumerate the number of CFUs expressing either YFP and CFP.

We imaged the colonies after 48 h of competition to visualize the distribution of the two different strains. *P. aeruginosa* natively produces secretions called siderophores that have a similar excitation and emission spectra that is similar to CFP[33,34]. While individual CFP-and YFP-labelled cells can easily be distinguished in the monolayer of subsurface colonies (Fig. 2g), surficial colonies are thicker, incubated for longer, and must be imaged with lower resolution objectives, making it difficult to distinguish the CFP-labelled cells from the secretions. To circumvent this problem, we imaged surficial colonies using a combination of brightfield and YFP fluorescence, such that regions with a larger fraction of CFP cells appeared darker grey in the merged brightfield/YFP images (Fig. 2a).

**Subsurface colony assay**

We prepared sub-surface colonies using a protocol similar to one previously described[35]. Briefly, a pad of 0.8% (w/v) LB agar was cut from a freshly poured plate and transferred to a glass slide. The agar pad was spotted with a 1 µl drop of bacterial culture adjusted to an optical density of $OD_{600} = 0.05$ (approximately 12,500 cells µl$^{-1}$), which was then allowed to dry until fully evaporated. The pad was then carefully inverted and placed into a Petri dish with a coverslip forming its base (175 µm glass thickness, MatTek), sandwiching the cells between agar and glass. By fully enclosing the agar pad, these dishes prevent evaporation and agar shrinkage over the course of the experiment. We found it was essential to use freshly poured agar to ensure consistency between experiments. We note that the bacterial culture was spotted onto the side of the agar that was facing up when it was initially poured (i.e. the side that was exposed to air rather than the side against the plastic Petri dish).

We used a 0.8% concentration of agar because preliminary experiments with cells that lack pili (Δ*pilB*)[20] and flagella (Δ*flgK*)[36] showed this concentration was soft enough to allow pili-based motility, but hard enough to suppress flagella-based motility. Control experiments showed that Δ*flgK* mutants form subsurface colonies that expand at a faster rate than WT cells, verifying that the colony expansion observed in our assay is not driven by flagellar motility (Extended Data Fig. 1). This finding is consistent with previous work that shows flagella mutants perform twitching motility at a faster rate[21], likely because flagella increase drag along the surface.

The evaporated droplet of cell culture used to initialize colonies contains a dense band of cells along its outer edge due to the "coffee ring" effect[37], while the interior of the evaporated droplet contains cells at much lower density. Analyses that investigate cells at low densities (Extended Data Figs. 3b, 5b, 6a) used images of cells in this interior region. Over time, cell division increases the density of cells within the interior region and the colony front begins to expand outward. In our assays that quantify the dynamics of these expanding colonies (Figs. 1e, f, 2c-g, Extended Data Fig. 1, Supplementary Movies 3, 4, 8), we used the dense outer band of cells on the edge of the evaporated droplet as the reference point from which to measure the colony radius, $r_{SUB}$. Assays that measure movement of cells in the monolayer of colonies (Figs. 1g, 3, Extended Data Figs. 3a, 8, Supplementary Movies 2, 5) were performed in colonies that had reached the steady-state expansion regime (Fig. 1f).

For subsurface colonies containing only a single strain (Fig. 1c-f, Supplementary Movie 3), we inoculated cultures of three different unlabelled strains (WT, Δ*pilH* and Δ*pilB*) at different positions on a single agar pad and used the motorized stage on the Zeiss inverted microscope to move between them at each imaging timepoint. The outer high-density band of cells was used to provide a consistent reference point from which to measure the colony radius, $r_{SUB}$. Each colony was imaged using a tile of eight adjacent fields of view, the first of which was centred on the band of aggregated cells at the edge of the "coffee ring" noted above. Eight brightfield images were acquired for each colony every two minutes over a period of 11 h. These experiments used a relatively low magnification (20X) to minimize the number of fields of views needed, allowing us to acquire data for all three colonies simultaneously.

For subsurface colonies containing two different fluorescently labelled strains (Fig. 2c-g), we inoculated 1 μL of the mixed culture onto an agar pad. Because these liquid-grown, exponential phase cells had relatively weak YFP and CFP signals, colonies were incubated at room temperature for 2.5 h prior to imaging. Colonies were imaged with the Zeiss inverted microscope. The colony edge was imaged using a tile of 20 adjacent fields of view, the first of which was centred on the edge of the colony at the start of imaging. To obtain a spatial resolution sufficient to estimate both cell packing fraction and genotypic composition, we used a 63X objective. The colony was imaged at ten-minute intervals to avoid phototoxicity and bleaching over the course of the 8 h experiment.

To track the movement of individual cells in the monolayer of colonies (Figs. 1g, 3, Extended Data Figs. 3a, 8c, Supplementary Movies 2, 5), we spotted monocultures of WT and Δ*pilH* cells onto separate agar pads. These were incubated overnight at room temperature (16 h) to allow the colony expansion to reach the steady-state regime (Fig. 1f). We then acquired 150X magnification brightfield images of the monolayer using the Nikon inverted microscope at a framerate of one image per second, yielding sufficient temporal and spatial resolution for single-cell tracking.

We also quantified the movement and size of individual cells at small packing fractions (Extended Data Figs. 3b, 5b, 6a) using separate subsurface experiments. In the centre of the inoculation spots, cell packing fractions were approximately 50-fold smaller than observed in the monolayer of a subsurface colony at steady-state, so cell-cell interactions were negligible[18].

To ensure cells had sufficient time to adapt to the surface[38], we began imaging 3 h after inoculation. For these low-density experiments, we used the 63X objective and a framerate of one second on the Zeiss inverted microscope.

**Imaging rosette formation**

Quantifying the movement of both defects and individual cells during the process of rosette formation (Fig. 4d-f, Extended Data Fig. 10, Supplementary Movie 9) was exceptionally challenging, as it required imaging the monolayer at high spatial resolution (63X magnification, two frames per minute) at precisely the time and place that rosettes begin to form. It was difficult to estimate *a priori* where rosette formation would occur and thus where to place the Zeiss inverted microscope's relatively small field of view to capture these events.

To maximize our chances of success, we inoculated multiple Δ*pilH*-YFP/WT-CFP subsurface colonies with 10 µl of culture at a range of different optical densities ($OD_{600}$) in a 6-well glass-bottomed plate (175 µm glass thickness, MatTek). We then imaged the monolayer of each colony in turn, starting from the colony initiated with the highest starting $OD_{600}$. As rosettes form earlier in colonies inoculated at higher densities, this provided multiple opportunities to image the monolayer precisely before rosette formation began.

Initially, we attempted to take time-lapse images of rosette formation using fluorescent confocal microscopy so we could continuously follow how the two strains were distributed using their YFP and CFP labels. However, this bleached the cells and adversely affected their movement. So instead, we imaged the dynamics of rosette formation using brightfield microscopy for a period of one hour. We then immediately switched over to confocal imaging, which allowed us to quantify the distribution and orientation of the two different strains within the same rosette.

**Liquid culture competition assay**

To compare the growth rate of mutants (Extended Data Fig. 5a), we grew the different strains in liquid culture and estimated their fitness relative to a WT reference strain by counting CFUs. We mixed a CFP labelled WT reference strain with YFP labelled Δ*pilH*, Δ*pilB*, and WT test strains in a 1:1 ratio. Liquid cultures were started at $OD_{600} = 0.02$ and placed in a shaking incubator at 23°C, the same temperature used in the subsurface colony experiments. We counted the number of YFP and CFP expressing CFUs after $t = 0$, 210, and 420 mins of competition to calculate the relative fitness, *w*, of the YFP test strain compared to that of the CFP control:

$$w = \frac{\ln(C_Y(t)/C_Y(0))}{\ln(C_C(t)/C_C(0))}, \qquad (2)$$

where $C_Y(0)$ and $C_C(0)$ are respectively the numbers of YFP and CFP cells measured at the beginning of the competition, and $C_Y(t)$ and $C_C(t)$ are the numbers of YFP and CFP cells at time $t$.

**Measurement of cell length in liquid cultures**

To measure the lengths of cells in liquid culture (Extended Data Fig. 6a), we combined exponentially growing cultures of Δ*pilH*-YFP and WT-CFP at a 1:1 ratio. These were fixed with 3% paraformaldehyde and then diluted in phosphate buffered saline (PBS, Fisher

Scientific) in 96-well plates with optical bottoms (Nunc brand, Thermo Scientific). We then centrifuged plates to ensure cells lay flat against the optical bottoms of the wells and imaged them using brightfield, YFP and CFP channels at 63X magnification. Cell lengths were then measured using our FAST software (see below).

## Analysis of Experimental Data

### Measuring subsurface colony expansion rates

We developed an image analysis pipeline to measure the characteristics of colonies as they spread across the surface. First, we use Fiji, an open source image analysis software[39], to perform background subtraction and to normalize the local contrast of the tile of brightfield images. The resulting images are then stitched together, generating a single image for each colony at each timepoint (Supplementary Movie 3). Next, we use a custom Matlab (MathWorks) script to identify the leading edge of the colony in the stitched images. Because the edge of the colony can undulate (see finger-like protrusions in Supplementary Movie 3), we divide the image into ten strips across its width and locate the position of colony edge in each by measuring where the image intensity rapidly changes. We then define $r_{SUB}$ as the median of these ten measurements. The colony expansion rate, $\frac{dr_{SUB}}{dt}$, is calculated by taking the gradient of $r_{SUB}$ and applying a smoothing filter to reduce high frequency noise.

### Defining the "front" and "homeland"

In the experiments where we competed fluorescently labelled WT and $\Delta pilH$ cells in subsurface colonies (Fig. 2c-g), we quantified the genotypic composition and cell packing fraction in both the "homeland" and the "front".

The homeland is defined as the region between $r_{SUB}$ = -10 µm and -60 µm at $t = 0$ (Fig. 2c, Supplementary Movie 4), (i.e., inside the evaporated drop of bacterial culture that was used to inoculate the colony). The position of the homeland remains fixed over the course of the experiment.

The front moves as the colony expands and is defined as the region that extends from the leading edge of the colony to 50 µm behind the leading edge. To account for undulations in the leading edge of the colony, we again subdivided the width of the image into ten strips, which can translate independently from one another (Supplementary Movie 4). This allowed us to more accurately measure the properties of cells at the periphery of the colony and avoid including the virgin agar beyond the leading edge in our automated analyses.

### Areal packing fraction measurements

To quantify how tightly packed cells were in the homeland and front regions, we calculated the areal packing fraction, which is defined as the proportion of the two-dimensional surface covered by cells. We note that the same number of rod-shaped *P. aeruginosa* cells can generate different areal packing fractions depending on both their orientations relative to the surface and if they are piled on top of one another. For example, after the monolayer of a colony becomes confluent (i.e. densely packed), the areal packing fraction may plateau even as the total number of cells per unit area continues to increase as the rod-shaped cells are realigned perpendicular to the surface.

The process used to calculate packing fraction is outlined in Extended Data Fig. 4: following background subtraction and contrast normalization, the brightfield image (Extended Data Fig. 4a) is segmented into a binary black and white image using a global intensity threshold (Extended Data Fig. 4b). This binary image contains dark ridges between densely-packed cells. Although useful for isolating individual cells, these ridges inaccurately suggest the existence of a space between touching cells. To remove them, morphological closure is next applied to the segmented image to generate a more accurate representation of the coverage of cells within the field of view (Extended Data Fig. 4c). The packing fraction is then calculated as the fraction of white pixels within this binary "coverage" image.

**Strain composition measurements**

To quantify changes of strain composition in both the homeland and front of subsurface colonies, we calculated the Δ*pilH* to total cell ratio (Δ*pilH* cells/total cells). We begin by stitching the tiles of YFP images (showing Δ*pilH* cells), CFP images (showing WT cells), and brightfield images (showing all cells) to form a single continuous image at each time point. The brightfield image is then segmented using a global intensity threshold (as during estimation of areal packing fraction, Extended Data Fig. 4b), allowing us to distinguish which pixels correspond to cells and which pixels correspond to unoccupied space. To discern whether the former corresponds to either a Δ*pilH* or WT cell, we calculate the YFP to CFP intensity ratio at each pixel. This helps to reduce the effect of systematic variations in fluorescence intensity (e.g., those caused by an uneven excitation field). The median value of the YFP to CFP ratio for pixels without cells is then used as a threshold for the pixels containing cells: pixels containing cells for which the YFP to CFP ratio is greater than this threshold are assigned as belonging to Δ*pilH* cells and pixels for which the ratio is below this threshold are assigned as belonging to WT cells. Our results are always verified by visual inspection; examples of the output of this process are shown in Fig. 2g. To calculate "Δ*pilH* cells/total cells" at each timepoint, the total number of pixels assigned to Δ*pilH* cells is divided by the total number of pixels belonging to both the WT and Δ*pilH* cells.

Shortly after the monolayer becomes confluent, the monolayer becomes three-dimensional and cells start to pile on top of one another. Once this occurs a given pixel in our epifluorescence images may contain fluorescence from more than one cell, making it difficult to distinguish the distributions of the two strains. Therefore, once the monolayer becomes confluent, we cease monitoring the composition of cell populations in the homeland.

**Single-cell tracking**

We attempted to use existing software packages to track movement of individual cells within the monolayer of our subsurface colony experiments. As a single monolayer image can contain more than 10,000 cells, we found existing software packages (designed for tracking objects at lower density) were prohibitively slow and/or were incapable of correctly segmenting individual cells when they are tightly packed together. Cell segmentation in this context is particularly challenging because neighbouring cells are often separated by a very subtle change in image intensity.

To overcome these problems, we developed a new Matlab-based tracking platform named FAST (Feature-Assisted Segmenter/Tracker). In brief, FAST uses a standard tracking by detection framework[40]. Firstly, individual cells are isolated from their neighbours using a sequence of segmentation routines. Next, we measure the "features" of each cell within each frame (including cell position, orientation, morphology, and fluorescence intensity). Finally, we use these features to follow individual cells between frames using an algorithm that automatically "trains" itself using machine learning to optimize tracking based on the available feature information.

In the segmentation stage, we use brightfield images to identify individual cells in the monolayer. Our software uses a combination of automated ridge detection[41], topographical watershed[42] and intensity thresholding to generate black and white binary images of cells that are not connected to one another. Using this binary image as a mask, we then extract each cell's "features" including its position, orientation, length, and width from the original brightfield image.

Tracking is achieved via a two-stage algorithm. In the first stage, a low-fidelity nearest-neighbour tracking algorithm is used to generate a set of putative links between objects in consecutive frames. The subset of links with the smallest corresponding frame-frame object displacements is then classified as correct, typically forming around half of the total putative links. This subset forms the training dataset. Statistical parameters can now be extracted from this collection of links, allowing the robustness of each feature as a marker of object identity to be measured. In the second stage of the tracking algorithm, these measurements are used to dynamically adjust the weighting of each feature such that unreliable features have a reduced weighting compared to more reliable features. Tracking is then repeated, using these reweighted features as inputs. This approach allowed us to obtain extremely large tracking datasets, for example yielding a total of 161,769 cell trajectories for the WT monolayer.

We note that, while FAST has primarily been developed for tracking of single cells in dense monolayers, its capabilities can also be leveraged for other datasets. For example, FAST can also be used to track topological defects through time and space (see next section).

**Detection and analysis of topological defects in experiments and the SPR model**

Comets and trefoils occur at singularities in cell orientation. Our automated approach for locating these singularities is similar to that described in[43]: in the first stage, we use the OrientationJ plugin for Fiji to measure the local orientation of cells in images using the tensor method[44]. Experimental monolayer images can be run directly through OrientationJ. For consistency, we also use the same defect analysis pipeline to analyse the output of our 2D SPR model: a timeseries of images are created by drawing rods as grey ellipses on a white background. These images are then processed using OrientationJ. To facilitate direct comparison between experimental and simulated systems, we set the size of the structure tensor window, which defines the spatial scale over which the orientation field is calculated, to a length equivalent to two cell/rod widths. This process yields the orientation of cells, $\theta = [-\pi/2, \pi/2)$ at each pixel in the input image (Fig. 3b).

The location of defects in the orientation fields are detected using a Matlab script that employs a discretized version of the standard path integral definition of a topological defect[11]. The topological charge, $n$, of each pixel is calculated as:

$$n = \frac{1}{2\pi} \sum_{i=2}^{9} [\theta_i - \theta_{i-1}], \qquad (3)$$

where $\theta_i$ is the cell orientation field at each of the 8 neighbouring pixels, where $\theta_9 = \theta_1$ so that the neighbouring pixels form continuous path. By definition, we order $\theta_1, \theta_2, \ldots \theta_9$ so they are ordered sequentially in an anticlockwise direction. This calculation is repeated for each pixel in the orientation field, except for those grid points along the edge of the field of view that are missing neighbors. Defect cores are detected as positions with non-zero values of $n$, with $n = +\frac{1}{2}$ at the location of comets, $n = -\frac{1}{2}$ at the location of trefoils, and $n = 0$ everywhere else.

Having located the defects, we next determine their orientations. To do this, we first define a set of positions $r$ that form an approximately circular path 5 pixels from the defect core. We then calculate the angle $\varphi(r)$ from the defect core to each of these positions and compare it to the value of the orientation field $\theta(\boldsymbol{r})$ at that position. The defect orientation is defined to be the value of $\varphi(\boldsymbol{r})$ at which the difference between $\varphi(\boldsymbol{r})$ and $\theta(\boldsymbol{r})$ is smallest[45].

We next use FAST to track the movement of defects using their position, orientation and topological charge as "features" (see above). We omit defects from our analyses that are present for fewer than five timepoints, as these are less reliable. This analysis yielded a total set of 1344 trefoil trajectories and 1382 comet trajectories.

The root mean square displacement (RMSD) of tracked trefoils and comets is calculated using:

$$R(\tau) = \sqrt{\langle \left(x(t+\tau) - x(t)\right)^2 + \left(y(t+\tau) - y(t)\right)^2 \rangle}. \quad (4)$$

where $\tau$ is the lag time, $(x, y)$ is the position of the defect, and $\langle \cdot \rangle$ denotes an ensemble average across all times $t$ and across all defect trajectories.

True to its name, pili-based "twitching" motility is jerky and highly unsteady, owing to the stochastic retraction and detachment of individual pili[7,46]. Obtaining a reliable measure of cell movement around defects thus required averaging of data across a large number of defects and cell trajectories so that the stochastic component of each cell's movement was averaged out.

To accomplish this, we first tracked the movement of defects and cells independently from one another. Next, we transformed the coordinate system of each cell trajectory so that its origin and orientation was measured relative to the centre and orientation of a nearby defect. This allowed us to combine cell trajectories collected from around a large number of comet and trefoil defects. After the cell trajectories were aligned within one another in the same reference frame, we averaged the cell velocity in a two-dimensional array of bins. The size of each bin was 3.2 µm × 3.2 µm. All velocity measurements were made with respect to the laboratory reference frame, not the reference frame of the defect.

We calculate the flowfield around defects in the SPR model in the same way as for the experiments, though cell trajectories are obtained directly from the model output, rather than from the FAST tracking software. To facilitate direct comparison between the non-dimensional SPR model and experiments, we normalized the flowfields around defects by dividing them by average speed of all cells within the simulation or field of view, respectively.

**PIV analysis of collective motility during rosette development**

Following the movement of individual cells during rosette formation was exceedingly difficult once they had reoriented perpendicular to the surface. Instead, we characterized the collective movement of cells during rosette formation using particle image velocimetry (PIV). This technique measures cell movement at a more coarse-grained level than single-cell tracking and does not require the segmentation of individual cells.

We performed PIV analysis on the timeseries of 63X magnification images recorded during rosette formation. Images were pre-processed using contrast normalization and manually stabilized to remove thermally-induced drift in the *xy*-plane. The resulting images were analysed using PIVlab, an open-source Matlab-based software[47]. We filtered our results using PIVlab's built-in tools to remove spurious measurements. Specifically, velocity vectors that exceeded 0.6 µm min$^{-1}$ were removed and replaced with velocities interpolated from surrounding neighbours, which helped to reduce noise. The resulting measurements of instantaneous velocity were then averaged over the entire one-hour period to obtain the mean movement of cells during the entire process of rosette formation (Fig. 4e, Extended Data Fig. 10).

**Modelling**

Motivation for individual-based modelling approach

Nematic systems have been modelled with a variety of different individual-based and continuum techniques[48–50]. When choosing which modelling framework to use to investigate our experimental system, we considered the following factors:

- First, we need a modelling framework that can capture cells in both the fully two-dimensional configuration that occurs at the beginning of our experiments, and during rosette formation where cells rotate out of the plane.

- Second, the cells in our experiment propel themselves using pili, which attach to a surface and retract to pull the body along[8]. This form of motility is intrinsically different to swimming motility, where cells propel themselves by exerting a force upon the fluid that is then balanced by hydrodynamic drag[51]. Hydrodynamic screening imposed by the nearby no-slip surface will render our experimental system "dry"[18,49,52], i.e. the hydrodynamic flow fields generated by cells are negligible and neighbouring cells align with each other purely via steric interactions.

- Third, rod-shaped *P. aeruginosa* cells are rigid. Unlike some bacterial species, whose bodies are flexible (e.g. *Myxococcus xanthus*[53]), *P. aeruginosa* cells maintain their straight morphologies even when packed together (Supplementary Movie 5).

- Fourth, we observe that isolated Δ*pilH* and WT cells move at different speeds (Extended Data Fig. 3b) and have slightly different aspect ratios (see Supplementary Notes, Extended Data Fig. S6a). We therefore require a model that allows cells with these differing properties to be mixed together in a single simulation.

With these considerations in mind, we decided that an individual based, Self-Propelled Rod (SPR) model[18,54] would allow us to best capture the processes occurring in our experiments. This approach models each cell as a rigid chain of Yukawa segments, with segments of different rods repelling one another with a Coulomb-like point potential. This repulsive interaction prevents rods from overlapping, simulating the steric interactions that occur between neighbouring cells. Varying the number of Yukawa segments in each rod allows us to simulate cells with different aspect ratios, whilst varying the propulsive force exerted by each rod allows us to simulate cells that move at different speeds.

**Self-propelled rod (SPR) models**

We initially implemented a version of the SPR model in which the orientation of rods was confined to a two-dimensional plane. This allowed us to simulate monolayers of cells prior to rosette development and facilitated direct comparison with a two-dimensional continuum model of active nematics (see section titled "Continuum model of active nematics"). To simulate the rosette formation, we subsequently extended our SPR model to allow rods to rotate out of the plane (see section titled "A three-dimensional SPR model to simulate rosette development").

**Two-dimensional SPR model**
Model Formulation

Following the governing equations presented in[54], we modelled cells as rigid rods composed of a fixed number of equally spaced Yukawa segments. Each rod is denoted by the index $\alpha$. A rod's centroid is given by $\boldsymbol{r}_\alpha = (x_\alpha, y_\alpha)$ and its orientation is given by the unit vector, $\hat{\boldsymbol{u}}_\alpha$, which, when rods are confined to two dimensions, can be described by a single angle, $\theta_\alpha$. We define $l_\alpha$ as the length of $\alpha$, $n_\alpha$ as the number of Yukawa segments in $\alpha$, and $\lambda$ as the screening length, the characteristic length scale over which the repulsive interaction between two segments decays. The characteristic aspect ratio of rod $\alpha$ is then given by $a_\alpha = l_\alpha/\lambda$.

The interaction potential, $U_{\alpha\beta}$, between two rods (denoted by subscripts $\alpha$ and $\beta$ respectively) is the sum of the interaction between each of their respective Yukawa segments:

$$U_{\alpha\beta} = \frac{U_0}{n_\alpha n_\beta} \sum_{i=1}^{n_\alpha} \sum_{j=1}^{n_\beta} \frac{e^{-r_{\alpha\beta}^{ij}/\lambda}}{r_{\alpha\beta}^{ij}}, \tag{5}$$

where $r_{\alpha\beta}^{ij} = \left(\left(x_\alpha^i - x_\beta^j\right)^2 + \left(y_\alpha^i - y_\beta^j\right)^2\right)^{\frac{1}{2}}$, is the Euclidian distance between the segments $i$ and $j$ in rods $\alpha$ and $\beta$ respectively, and $U_0$ is the potential amplitude. The total interaction potential for each rod, $U_\alpha$, is then equal to the sum of all of the interactions between $\alpha$ and all of the other rods in the simulation.

The equations of motion that describe the translation and rotation of each rod, are respectively:

$$\boldsymbol{f}_T \cdot \frac{\partial \boldsymbol{r}_\alpha}{\partial t} = -\frac{\partial U_\alpha}{\partial \boldsymbol{r}_\alpha} + F\hat{\boldsymbol{u}}_\alpha, \tag{6a}$$

$$f_\theta \frac{\partial \theta_\alpha}{\partial t} = -\frac{\partial U_\alpha}{\partial \theta_\alpha}, \tag{6b}$$

where $\boldsymbol{f}_T$ is the translational friction tensor, $f_\theta$ is the rotational friction constant and $F$ is the size of the force exerted by a rod along its axis. We use the formulation presented in[54] to calculate $\boldsymbol{f}_T$ and $f_\theta$, which are in turn a function of the rod aspect ratio, $a_\alpha$, and the Stokesian friction coefficient, $f_0$.

Simulations

Our SPR model was implemented in Matlab. Following the approach of[18] we set $\lambda = 1$, $F = 1$ and $f_0 = 1$. The dynamics of this SPR model has previously been shown to only weakly depend upon the potential amplitude $U_0$, provided its value is sufficiently large to prevent rod-rod crossing. We used $U_0 = 250$, which falls within this range[18]. Our simulations were initialized with all rods in uniformly spaced rows and with their orientations aligned with the *y*-axis. Each rod was randomly assigned a movement direction (i.e. half of the rods exerted a force along +*y* and half along −*y*). After an initial transient, this arrangement of rods quickly gives rise to a system that exhibits local nematic ordering that is directed in random directions. We ensured that all simulations reached a statistical steady state before using their output in our analyses.

We calculate the packing fraction of rods, $\rho$, in our simulations as the total area of all rods divided by area of the computational domain. In keeping with previous studies[18,54], we model the rods as stadia (i.e. as rectangles with semi-circular caps on either end) of length $l_\alpha$ and width $\lambda$. This yields:

$$\rho = \frac{1}{A} \sum_{\alpha=1}^{N} \left[ \lambda(l_\alpha - \lambda) + \frac{\pi \lambda^2}{4} \right], \tag{7}$$

where $N$ is the number of rods ($10^3 < N < 10^4$) and $A$ is the total area of the computational domain. Our simulations correspond to $\rho = 0.25$, which for $F = 1$ produced collective behavior similar to that which was observed in the WT monolayer. We note that the characteristic length scale of repulsion of the Yukawa segments, $\lambda$, is an approximation of rod "width", so a domain containing a very tightly packed assemblage of rods could potentially yield $\rho > 1$.

We integrated the governing equations using the midpoint method and periodic boundary conditions. The timestep used ($\Delta t = 0.2$) provided numerically stable results for all sets of simulation parameters tested. To ensure that our results were independent of the size of our domain, we fixed the system density at $\rho = 0.25$ and varied the number of rods, $N$. We found the rod speed and verticalization were independent of the number of rods in our simulation provided that $N > 10^3$.

Simulating the flow around topological defects using the 2D SPR model

We used our two-dimensional SPR model to predict the flow of cells around comets and trefoils (Fig. 3d). We set $N = 5000$ and chose an aspect ratio $a = 4$ to match the morphology of WT cells in high-density monolayers (Extended Data Fig. 6a). To ensure that the experiments and simulations were analysed in the same way, we used the simulations to generate an image of rod positions at each timepoint and then used OrientationJ to detect topological defects. However, we obtained rod trajectories directly from the simulation output, rather than via cell tracking software. To compare our experiments with simulations, we normalized the dimensions of the former by cell width and the latter by $\lambda$, similar to previous studies[18].

**Three-dimensional SPR model**

Simulation of rosette formation required that the SPR model be extended into the third dimension to allow rods to reorient vertically. To achieve this, we introduce two new variables into our model: firstly, we add the coordinate $\boldsymbol{r}_\alpha = (x_\alpha, y_\alpha, z_\alpha)$ to our representation of the rod centroid $\boldsymbol{r}_\alpha$. Secondly, we allow rods to alter their angle $\phi_\alpha$ relative to the $xy$-plane.

To define the equation of motion for $\phi_\alpha$, we begin by calculating the torque about the centroid of rod $\alpha$ that acts to rotate it out of the $xy$-plane. This torque is generated by steric interactions with surrounding rods and is proportional to the potential gradient $-\frac{\partial U_\alpha}{\partial \phi_\alpha}$. Note that this is equivalent to the term $-\frac{\partial U_\alpha}{\partial \theta_\alpha}$ in equation 6b, which similarly drives the rotation of rods in the $xy$-plane.

To rotate out of plane, cells must overcome the stabilizing effect of the adhesive secretions that act to glue cells to the surfaces (i.e., extracellular polymeric secretions[55]). In the subsurface assay the overlaying agar also acts to keep the long axis of cells flat against the glass surface[29]. To simulate the resistance to rotation exerted by both cell secretions and agar, we incorporate a physically-based stabilizing torque into our three-dimensional model that acts to keep rods flat against the surface.

Both polymeric secretions and agar have been shown to generate a linearly elastic restoring force when they are deformed on length scales equivalent to that of bacterial cells[56,57]. To simulate the influence of this elastic response on the dynamics of $\phi_\alpha$, we assume that the substrate generates a force proportional to the upwards displacement of the tilting rod. This will generate a restoring force equal to $kl_\alpha \sin \phi_\alpha$ at the rod's tip, where $k$ is the elastic modulus of the substrate. Transforming this into a torque acting on the centroid of rod $\alpha$ and adding the torque imposed by the surrounding rods, we arrive at the final equations of motion for the rod, $\alpha$:

$$\boldsymbol{f}_T \cdot \frac{\partial \boldsymbol{r}_\alpha}{\partial t} = -\frac{\partial U_\alpha}{\partial \boldsymbol{r}_\alpha} + F\hat{\boldsymbol{u}}_\alpha, \tag{8a}$$

$$f_\theta \frac{\partial \theta_\alpha}{\partial t} = -\frac{\partial U_\alpha}{\partial \theta_\alpha}, \tag{8b}$$

$$f_\phi \frac{\partial \phi_\alpha}{\partial t} = -\frac{\partial U_\alpha}{\partial \phi_\alpha} + \frac{kl_\alpha^2}{2}\cos \phi_\alpha \sin \phi_\alpha. \tag{8c}$$

The friction constant $f_\phi$ provides viscous damping for reorientations in the $\phi_\alpha$ direction; we assume that viscosity is equal in all directions, i.e. that $f_\phi = f_\theta$. This new model therefore contains one new parameter, $k$, the elastic modulus of the substrate. In practice, we set $k = 0.6$, ensuring the transition to 3D rosette formation occurs over the range of self-propulsion forces $1 < F < 2$. Similar behaviors are observed for other values of $k$, except the transition occurs at other values of $F$. Finally, we note our simulations enforced $\frac{\partial z_\alpha}{\partial t} = 0$, which allowed the rods to tilt out of the *xy*-plane, but maintained their centroids at $z = 0$.

3D SPR model of comet collision

To conceptualize how the collision of +½ defects (comets) gives rise to rosettes, we initialized three-dimensional SPR simulations with two comets directed towards one another (Fig. 4a, Supplementary Movie 6). Simulations were initialized as a 12 by 12 lattice of evenly spaced rods, but in contrast to the other simulations described in this section (where rods were randomly initialized along either the +*y* or -*y* direction), rod orientations, $\theta_\alpha$, were initialized using a pre-defined director field consisting of two comet defects pointing at each other (Fig. 4a). Initial values of $\phi_\alpha$ were drawn from a normal distribution with zero mean and a standard deviation of 0.5° degrees, which allowed the rods to escape the unstable stationary point at $\phi_\alpha = 0$ (equation 8*c*). Simulations with different self-propulsion forces, $F$, were run using identical starting configurations. All rods within a given simulation had the same value of $F$.

3D SPR simulations initialized with random rod orientations

To investigate how changes in both cell aspect ratio, *a*, and force generation, *F*, affect rosette formation in more disordered monolayers, we performed separate simulations at different values of *F* and *a* (Extended Data Figs. 6b, c, 9, Supplementary Movie 7). All rods in a given simulation possessed identical parameters. $N = 1600$ rods were initialized on a 2D lattice in the *xy*-plane, with a small perturbation to their tilt $\phi_\alpha$ as noted in the previous section. This system was allowed to evolve for 300 time steps with $\frac{\partial \phi_\alpha}{\partial t} = 0$, which confined the rods' orientation to the *xy*-plane as they reached a random 2D nematic configuration. We then allowed rods to tilt out of plane and simulated the resulting 3D dynamics for another 2000 time steps. Both the mean rod speed and mean proportion of rods in a vertical orientation were then calculated over the final 500 time steps. While a rod in a perfectly vertical orientation corresponds to $\phi_\alpha = 90°$, in practice the orientation of rods within rosettes fluctuate around this value, so we considered a rod to be "vertically oriented" if $\phi_\alpha > 85°$. Three simulations were performed for each set of parameter values tested, randomizing the initial configuration of rods for each repeat.

3D SPR simulations to explore the interactions between different genotypes

To simulate the interaction of two different genotypes that each generate a different propulsive force, we performed simulations using a similar methodology to that described in the previous section. However, half of the rods propelled themselves with $F_1 = F_v = 1.5$ and the other half with variable $F_2$ (Fig. 4b, c). To explore how changes in cell length affects the interaction between WT and $\Delta pilH$ cells, we also performed simulations where we mixed rods that differ in both their force generation and aspect ratio. Half the rods had $a_1 = 4$ and $F_1 = F_v = 1.5$, while the other half had $a_2 = 5$ and exerted a force $F_2$ that was varied in different simulations (Extended Data Fig. 6d, e).

**Continuum model of active nematics**

In addition to our SPR model, we also compared our experimental results with an analytical prediction of the flow of cells around topological defects using a continuum description of active nematics[11,12,58–60]. We used governing equations to describe the coarse-grained velocity $\boldsymbol{u} = (u_r, u_\theta)$ and director fields $\boldsymbol{n} = (n_r, n_\theta)$, the latter representing the orientation of bacteria relative to the center of a defect with charge, *m*. We considered both trefoils (*m* = −1/2) and comets (*m* = +1/2), with each defect occurring at the centre of a circular domain with radius *R* and a no-slip boundary. This confinement models a defect's interaction with neighbouring topological defects, which act to screen the flow field about it. Despite of the polarity of the self-propelled bacteria, the use of a nematic director to describe their orientation is motivated by the emergence of half-integer, nematic, topological defects in the experiments and in the SPR model, which clearly indicate the existence of nematic symmetry within the monolayer. The emergence of such a nematic (apolar) symmetry of a system consisting of polar self-propelled rods is a well-known phenomenon in active matter systems[61–63].

To analytically calculate the flow field we make several simplifying assumptions, following the approach used in[11]. We assume that:

(*i*) the size of the defect core, *a*, is much smaller than the screening length scale, *R*, and that we can therefore neglect variations in the magnitude of the orientational order, (i.e., how strongly the cells are aligned at each point). In reality, the orientational order is zero at the defect core and continuously increases to a finite value, which depends on the thermodynamic properties of the system. Assuming that orientational order around a defect has a uniform magnitude, we can write the director field as:

$$\boldsymbol{n} = (n_r, n_\theta) = (\cos[(m-1)\theta], \sin[(m-1)\theta]), \qquad (9)$$

where *m* is −1/2 or +1/2.

(*ii*) We further assume that at steady state the flow field does not alter the director configuration around a defect, so that a separate equation for the evolution of $\boldsymbol{n}$ is not required.

(*iii*) Finally, we assume that the active stresses generated by bacteria dominate over passive elastic stresses that arise due to the orientational deformations around the defect. With this assumption, we can write the Stokes equation simply as a balance of the active force generation and the viscous dissipation

$$0 = -\boldsymbol{\nabla} p + \eta \boldsymbol{\nabla}^2 \boldsymbol{u} - \zeta \boldsymbol{\nabla} \cdot \left(\boldsymbol{n}\boldsymbol{n}^T - \frac{I}{2}\right), \qquad (10)$$

where $p$ is the pressure, $\eta$ is the viscosity, and $\zeta$ is the activity coefficient. The direction of flows around defects in our experiments (Fig. 3d) imply that the activity, $\zeta$, is positive, i.e. that our system is extensile. The last term on the right hand side represents the active forces generated by the bacteria and is obtained by coarse-graining the equal and opposite forces arising from self-propulsion and the resistance exerted by the substrate[58,63,64]. Using the form of the director around topological defects from Eq. (9), Eq. (10) can be solved in terms of Green's functions, to yield the velocity field around positive half-integer defects (comets):

$$u_r^{comet} = \left(\frac{\zeta R}{12\eta} - \frac{\zeta}{6\eta}r + \frac{\zeta}{12\eta R}r^2\right)\cos[\theta], \qquad (11a)$$

$$u_\theta^{comet} = \left(-\frac{\zeta R}{12\eta} + \frac{\zeta}{4\eta}r - \frac{\zeta}{4\eta R}r^2\right)\sin[\theta], \qquad (11b)$$

whilst for negative half-integer defects (trefoils) we have:

$$u_r^{trefoil} = \left(-\frac{\zeta}{10\eta}r + \frac{3\zeta}{20\eta R}r^2 - \frac{\zeta}{20\eta R^3}r^4\right)\cos[3\theta], \qquad (12a)$$

$$u_\theta^{trefoil} = \left(\frac{\zeta}{15\eta}r - \frac{3\zeta}{20\eta R}r^2 + \frac{\zeta}{12\eta R^3}r^4\right)\sin[3\theta]. \qquad (12b)$$

The details of the algebra are the same as[11] and therefore are omitted here.

**Data availability**

Data that support the findings of this study can be accessed at https://doi.org/10.15131/shef.data.12735251.v1.

**Code availability**

The FAST cell tracking package can be accessed at https://doi.org/10.5281/zenodo.3630641, with extensive documentation on its use and functionality available at https://mackdurham.group.shef.ac.uk/FAST_DokuWiki/dokuwiki. The Defector defect detection package is available at https://doi.org/10.5281/zenodo.3974873, while the colEDGE colony composition package can be accessed at https://doi.org/10.5281/zenodo.3974875.

**Supplementary References**

31. O'Toole, G. A. & Kolter, R. Flagellar and twitching motility are necessary for Pseudomonas aeruginosa biofilm development. *Mol. Microbiol.* **30**, 295–304 (1998).

32. Choi, K.-H. & Schweizer, H. P. mini-Tn7 insertion in bacteria with single attTn7 sites: example Pseudomonas aeruginosa. *Nat. Protoc.* **1**, 153–161 (2006).

33. Cox, C. D. & Graham, R. Isolation of an iron-binding compound from Pseudomonas aeruginosa. *J. Bacteriol.* **137**, 357–64 (1979).

34. Elliott, R. P. Some properties of pyoverdine, the water-soluble fluorescent pigment of the pseudomonads. *Appl. Microbiol.* **6**, 241–6 (1958).

35. Darzins, A. The pilG gene product, required for Pseudomonas aeruginosa pilus production and twitching motility, is homologous to the enteric, single-domain response regulator CheY. *J. Bacteriol.* **175**, 5934–44 (1993).

36. O'Toole, G. A. & Kolter, R. Initiation of biofilm formation in Pseudomonas fluorescens WCS365 proceeds via multiple, convergent signalling pathways: a genetic analysis. *Mol. Microbiol.* **28**, 449–461 (1998).

37. Deegan, R. D. *et al.* Capillary flow as the cause of ring stains from dried liquid drops. *Nature* **389**, 827–829 (1997).

38. Luo, Y. *et al.* A hierarchical cascade of second messengers regulates Pseudomonas aeruginosa surface behaviors. *MBio* **6**, e02456-14 (2015).

39. Schindelin, J. *et al.* Fiji: an open-source platform for biological-image analysis. *Nat. Methods* **9**, 676–682 (2012).

40. Li, K. *et al.* Cell population tracking and lineage construction with spatiotemporal context. *Med. Image Anal.* **12**, 546–566 (2008).

41. Lindeberg, T. Edge Detection and Ridge Detection with Automatic Scale Selection. *Int. J. Comput. Vis.* **30**, 117–156 (1998).

42. Meyer, F. Topographic distance and watershed lines. *Signal Processing* **38**, 113–125 (1994).

43. Kawaguchi, K., Kageyama, R. & Sano, M. Topological defects control collective dynamics in neural progenitor cell cultures. *Nature* **545**, 327–331 (2017).

44. Püspöki, Z., Storath, M., Sage, D. & Unser, M. Transforms and Operators for Directional Bioimage Analysis: A Survey. in *Focus on Bio-Image Informatics* (eds. Vos, W. H. De, Munck, S. & Timmermans, J.-P.) 69–93 (Springer, Cham, 2016). doi:10.1007/978-3-319-28549-8_3

45. Huterer, D. & Vachaspati, T. Distribution of singularities in the cosmic microwave background polarization. *Phys. Rev. D* **72**, 043004 (2005).

46. Jin, F., Conrad, J. C., Gibiansky, M. L. & Wong, G. C. L. Bacteria use type-IV pili to slingshot on surfaces. *PNAS* **108**, 12617–22 (2011).

47. Thielicke, W. & Stamhuis, E. J. PIVlab – Towards User-friendly, Affordable and Accurate Digital Particle Image Velocimetry in MATLAB. *J. Open Res. Softw.* **2**, e30 (2014).

48. DeCamp, S. J., Redner, G. S., Baskaran, A., Hagan, M. F. & Dogic, Z. Orientational order of motile defects in active nematics. *Nat. Mater.* **14**, 1110–1115 (2015).

49. Doostmohammadi, A., Adamer, M. F., Thampi, S. P. & Yeomans, J. M. Stabilization of active matter by flow-vortex lattices and defect ordering. *Nat. Commun.* **7**, 10557 (2015).

50. Shi, X. & Ma, Y. Topological structure dynamics revealing collective evolution in active nematics. *Nat. Commun.* **4**, 3013 (2013).

51. Baskaran, A. & Marchetti, M. C. Statistical mechanics and hydrodynamics of bacterial suspensions. *PNAS* **106**, 15567–72 (2009).

52. Drescher, K., Dunkel, J., Cisneros, L. H., Ganguly, S. & Goldstein, R. E. Fluid dynamics and noise in bacterial cell-cell and cell-surface scattering. *PNAS* **108**, 10940–5 (2011).

53. Harvey, C. W. *et al.* Study of elastic collisions of Myxococcus xanthus in swarms. *Phys. Biol.* **8**, 026016 (2011).

54. Wensink, H. H. & Löwen, H. Emergent states in dense systems of active rods: from swarming to turbulence. *J. Phys. Condens. Matter* **24**, 464130 (2012).

55. Zhao, K. *et al.* Psl trails guide exploration and microcolony formation in Pseudomonas aeruginosa biofilms. *Nature* **497**, 388–91 (2013).

56. Klapper, I., Rupp, C. J., Cargo, R., Purvedorj, B. & Stoodley, P. Viscoelastic fluid description of bacterial biofilm material properties. *Biotechnol. Bioeng.* **80**, 289–296 (2002).

57. Nayar, V. T., Weiland, J. D., Nelson, C. S. & Hodge, A. M. Elastic and viscoelastic characterization of agar. *J. Mech. Behav. Biomed. Mater.* **7**, 60–68 (2012).

58. Aditi Simha, R. & Ramaswamy, S. Hydrodynamic Fluctuations and Instabilities in Ordered Suspensions of Self-Propelled Particles. *PRL* **89**, 058101 (2002).

59. Marenduzzo, D., Orlandini, E. & Yeomans, J. M. Hydrodynamics and Rheology of Active Liquid Crystals: A Numerical Investigation. *PRL* **98**, 118102 (2007).

60. Marenduzzo, D., Orlandini, E., Cates, M. E. & Yeomans, J. M. Steady-state hydrodynamic instabilities of active liquid crystals: Hybrid lattice Boltzmann simulations. *Phys. Rev. E* **76**, 031921 (2007).

61. Kraikivski, P., Lipowsky, R. & Kierfeld, J. Enhanced Ordering of Interacting Filaments by Molecular Motors. *PRL* **96**, 258103 (2006).

62. Baskaran, A. & Cristina Marchetti, M. Nonequilibrium statistical mechanics of self-propelled hard rods. *J. Stat. Mech. Theory Exp.* **2010**, P04019 (2010).

63. Marchetti, M. C. *et al.* Hydrodynamics of soft active matter. *Rev. Mod. Phys.* **85**, 1143–1189 (2013).

64. Thampi, S. P., Golestanian, R. & Yeomans, J. M. Velocity Correlations in an Active Nematic. *PRL* **111**, 118101 (2013).

65. Barken, K. B. *et al.* Roles of type IV pili, flagellum-mediated motility and extracellular DNA in the formation of mature multicellular structures in Pseudomonas aeruginosa biofilms. *Environ. Microbiol.* **10**, 2331–43 (2008).

66. Fulcher, N. B., Holliday, P. M., Klem, E., Cann, M. J. & Wolfgang, M. C. The Pseudomonas aeruginosa Chp chemosensory system regulates intracellular cAMP levels by modulating adenylate cyclase activity. *Mol. Microbiol.* **76**, 889–904 (2010).

67. Persat, A., Inclan, Y. F., Engel, J. N., Stone, H. A. & Gitai, Z. Type IV pili

mechanochemically regulate virulence factors in Pseudomonas aeruginosa. *PNAS* **112**, 7563–8 (2015).

68. O'Toole, G. A. & Wong, G. C. Sensational biofilms: surface sensing in bacteria. *Curr. Opin. Microbiol.* **30**, 139–146 (2016).

69. Aris, R. & Humphrey, A. E. Dynamics of a chemostat in which two organisms compete for a common substrate. *Biotechnol. Bioeng.* **19**, 1375–1386 (1977).

70. Smith, W. P. J. *et al.* Cell morphology drives spatial patterning in microbial communities. *PNAS* **114**, E280–E286 (2017).

71. Stenhammar, J., Wittkowski, R., Marenduzzo, D. & Cates, M. E. Activity-Induced Phase Separation and Self-Assembly in Mixtures of Active and Passive Particles. *PRL* **114**, 018301 (2015).

72. Straley, J. P. Frank Elastic Constants of the Hard-Rod Liquid Crystal. *Phys. Rev. A* **8**, 2181–2183 (1973).

73. Priest, R. G. Theory of the Frank Elastic Constants of Nematic Liquid Crystals. *Phys. Rev. A* **7**, 720–729 (1973).

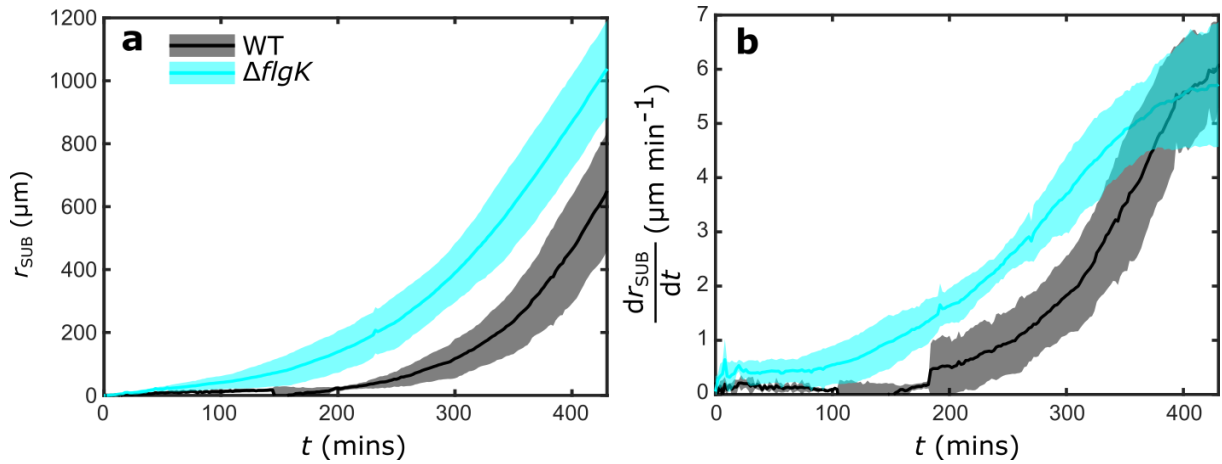

**Extended Data Fig. 1 | Flagella do not drive the expansion of subsurface colonies. a**, **b,** Measurements of the colony radius, $r_{SUB}$, and colony expansion rate, $\frac{dr_{SUB}}{dt}$, for both the flagellated wild-type (WT) and a non-flagellated mutant ($\Delta flgK$) strain. These experiments indicate that flagella actually hinder colony expansion. We speculate this is because flagella actively stick to surfaces[21], increasing the cells' resistance to movement. Shaded regions indicate the standard deviation about the mean for three separate experiments.

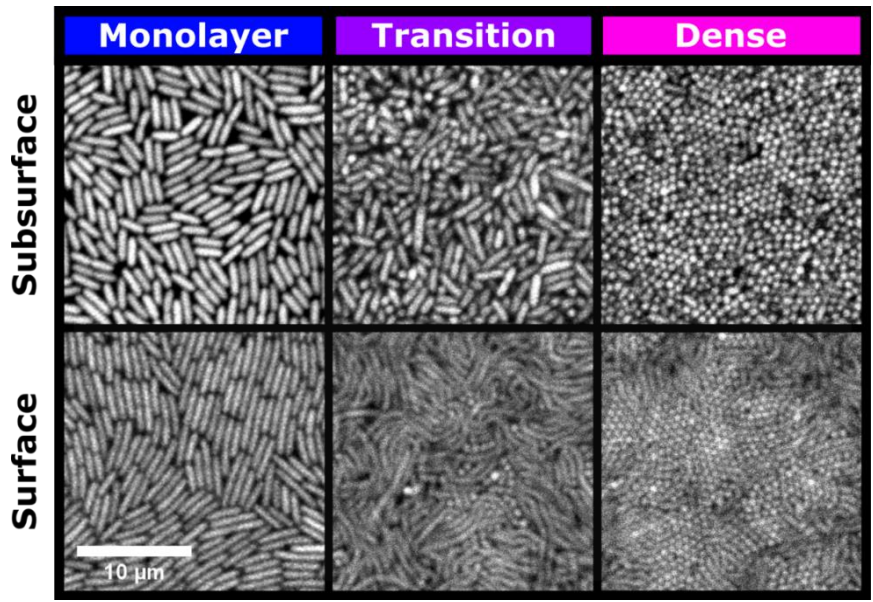

**Extended Data Fig. 2 | Cells exhibit similar orientations in surficial and sub-surface colonies.** Both surficial and subsurface colonies of YFP labelled WT cells were grown for 24 h at room temperature and then imaged using confocal microscopy (Methods). Both subsurficial (upper row) and surficial (lower row) colonies have a "monolayer" of cells lying flat against the surface at their periphery, a "dense" region where most cells are vertically standing up on end at their centre, and a "transition" region where some cells are standing up between the colony edge and centre (Fig. 1b-d). We note that subsurface colony images shown here are the same as those presented in Fig. 1d, but are reproduced here to facilitate direct comparison.

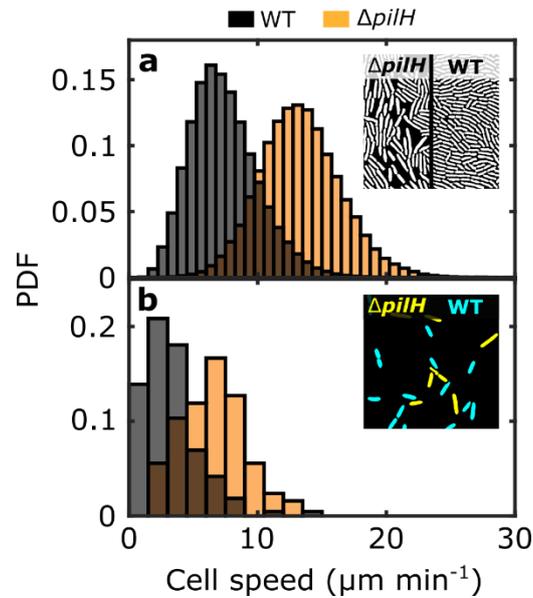

**Extended Data Fig. 3 | Δ*pilH* cells move faster than WT cells.** Cells in the monolayer of Δ*pilH* colonies move faster than those in WT colonies (Fig. 1g, reproduced in **a**). However, Δ*pilH* monolayers are also observed to have a smaller packing fraction than WT monolayers (**a**, inset). To test if the variation in cell density could confound our analyses, we also performed a separate experiment in which Δ*pilH* and WT cells were mixed together at low density. Separate fluorescent markers were used to distinguish strains (**b**, inset). This confirmed that Δ*pilH* cells move more quickly than WT cells when the two are at equal density ($p < 10^{-17}$, Mann-Whitney *U* test). All experiments shown were performed using the subsurface assay.

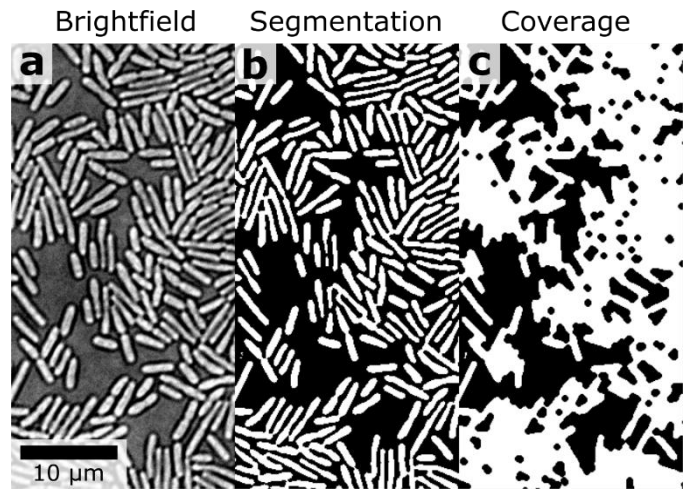

**Extended Data Fig. 4 | Automated calculation of areal packing fraction of cells in subsurface colonies.** A brightfield image (**a**, preprocessed as described in Methods) is first converted into a black and white binary image using a global intensity threshold (**b**). Next, we use morphological closure to remove the boundaries between cells that are touching one another, allowing us to identify the areas where cells fully cover the surface (**c**). The packing fraction is then calculated as the proportion of white pixels in the resulting "coverage" image (in this case, 0.60). See Methods for additional details.

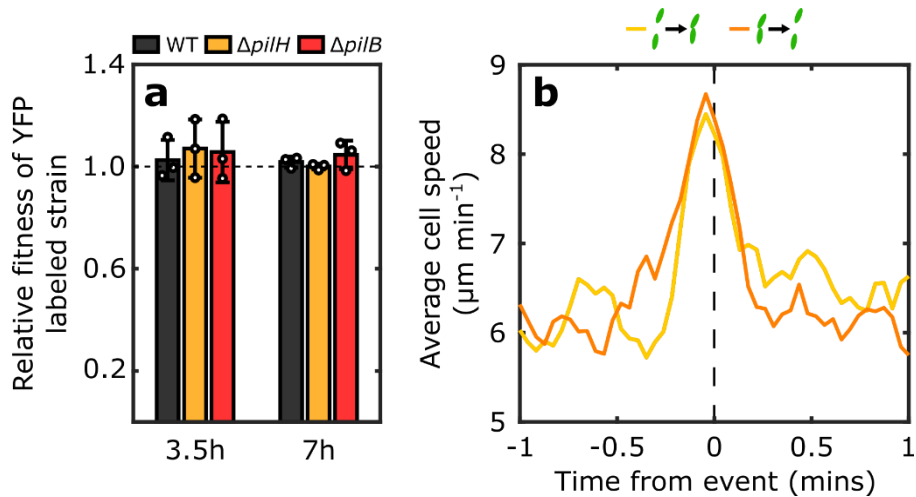

**Extended Data Fig. 5 | Δ*pilH* cells do not adhere to one another, and do not grow more slowly than WT cells. a**, Relative fitness of three YFP labelled test strains (WT, Δ*pilH* and Δ*pilB*) compared to a CFP labelled WT reference strain co-cultured in liquid culture. The relative fitness of each test strain was not significantly different from 1 at either 3.5 or 7 h post-inoculation ($p > 0.05$, one sample t-test, $n = 3$, Methods). Error bars indicate standard deviation of 3 replicates. **b**, We measured the mean speed of previously solitary Δ*pilH* cells as they came into contact with one another (light orange line, $n = 41$) and the mean speed of Δ*pilH* cells already in contact with one another as they moved apart (dark orange, $n = 47$). If cells actively adhered to each other, we would expect them to slow down after contacting one another and increase their speed after moving away from one another (Supplementary Notes). We find that cell speed peaks at $t = 0$, which corresponds to the time point at which cells either make or break contact. However, in both cases we observed that there was no appreciable change in cell speed after either event, indicating that Δ*pilH* cells do not adhere to each other.

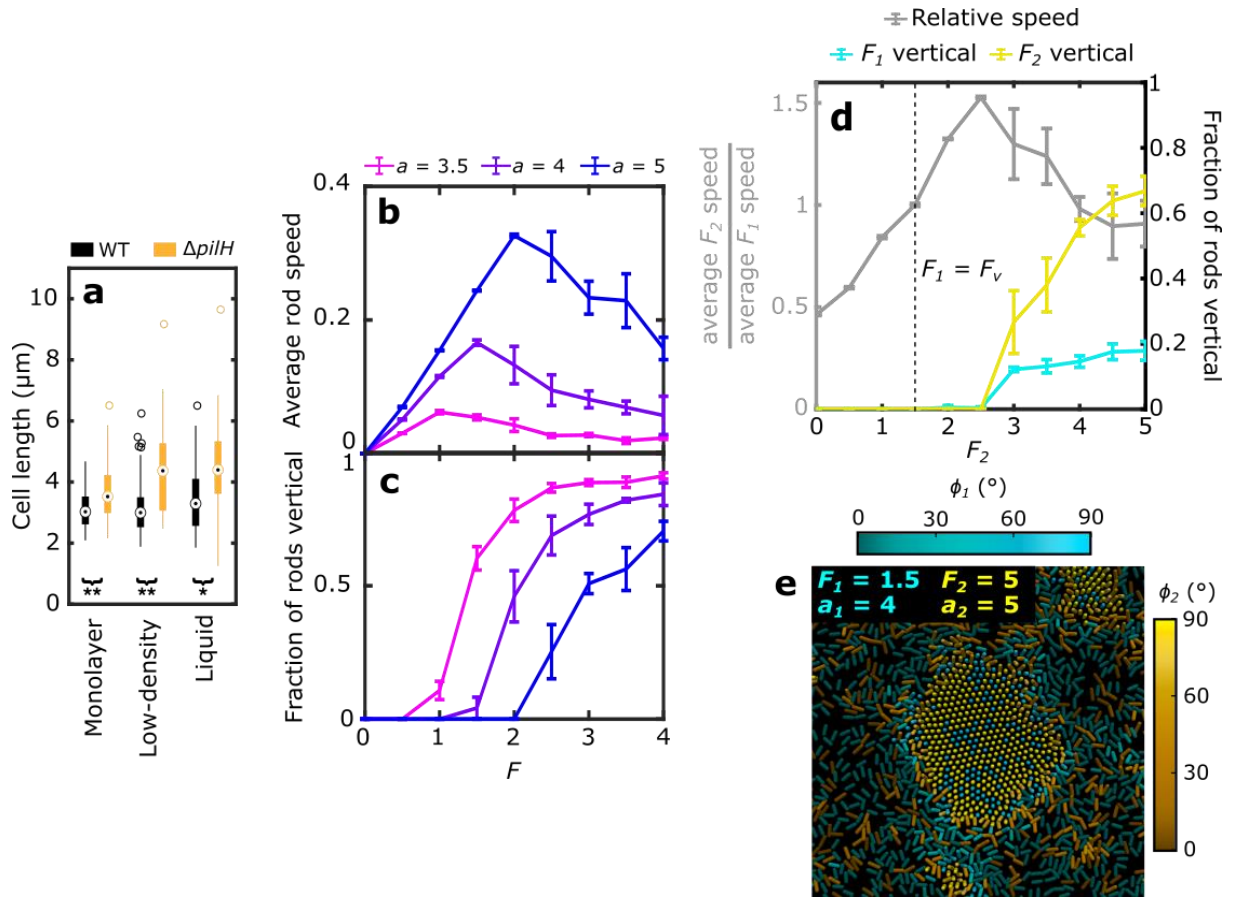

**Extended Data Fig. 6 | Δ*pilH* cells are longer than WT cells, which stabilizes them against verticalization but increases their representation in rosettes when mixed with a shorter genotype. a**, Boxplots of lengths of WT (black) and Δ*pilH* (orange) cells mixed together in a high-density subsurface colony ("Monolayer", WT $n = 223$, Δ*pilH* $n = 218$), a low-density subsurface colony ("Low-density", WT $n = 114$, Δ*pilH* $n = 84$), and in liquid culture at exponential phase ("Liquid", WT $n = 60$, Δ*pilH* $n = 34$) (Methods). The Δ*pilH* cells were significantly longer than WT cells in all three environments (* = $p < 10^{-3}$, ** = $p < 10^{-10}$, Mann-Whitney *U* test). Boxplots indicate median (central white rings), interquartile range (box limits), 1.5x interquartile range (whiskers) and outliers (individual circles). **b**, Average rod speed at steady state in 3D SPR monolayer simulations for rods with different propulsive forces, *F*, and rod aspect ratio, *a*. All rods in a given simulation are identical to one another. **c**, Proportion of rods oriented vertically at the end of simulations shown in **b**. **d**, Steady state velocity and verticalization measurements for simulations in which a "mutant" population of rods that are propelled by a variable force $F_2$ and with a fixed aspect ratio $a_2 = 5$ interacts with a "wild-type" population with $F_1 = F_v = 1.5$ (fixed), $a_1 = 4$ (fixed). These simulations are similar to the ones shown in Fig. 4b, except the two populations of rods also have different aspect ratios. Error bars in **b**-**d** indicate the standard deviation of three separate simulations, each with a different random initial configuration. **e**, A rosette spontaneously generated in co-culture simulation with parameters $F_1 = F_v = 1.5$, $a_1 = 4$, $F_2 = 5$ and $a_2 = 5$ illustrates how the longer length of the mutant enhances its representation in rosettes (compare with Fig. 4c and see Supplementary Notes).

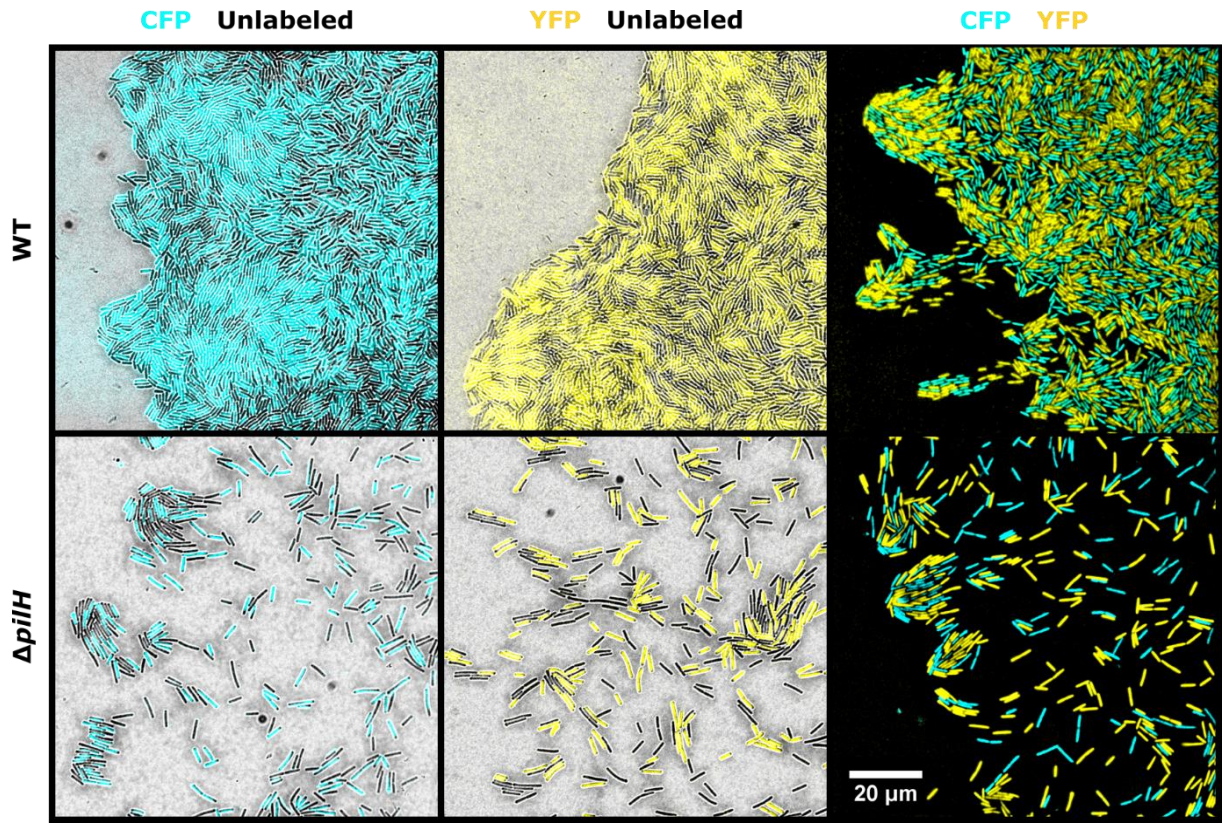

**Extended Data Fig. 7 | Fluorescent labelling does not impact the ability of *P. aeruginosa* strains to compete in colonies.** The leading edge of six different subsurface colonies inoculated with equal fractions of CFP-labelled and unlabelled strains (left), YFP-labelled and unlabelled strains (middle), and CFP-labelled and YFP-labelled strains (right) after 16 h of incubation at room temperature. WT and Δ*pilH* colonies are shown on the upper and lower rows respectively. In all six colonies, near equal proportions of each cell type are present at the colony's leading edge indicating that expression of a fluorescent label has a negligible impact on each strain's competitive ability. Unlabelled strains are imaged using brightfield and appear grey.

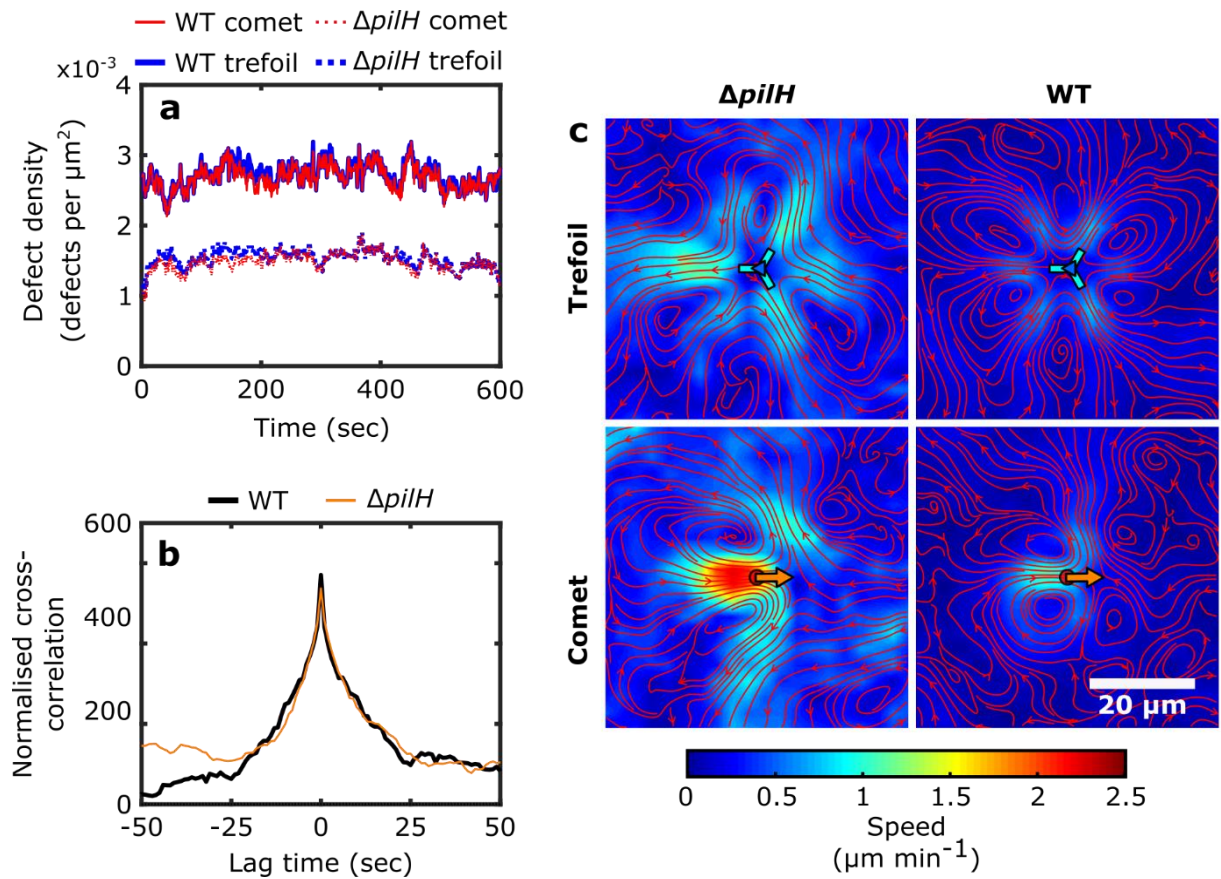

**Extended Data Fig. 8 | Automated analysis of defects reveals differences between the collective motility of WT and Δ*pilH* monolayers. a**, We used automated defect detection to count the number of comets/trefoils and normalized these by the area of the field of view (Methods). Averaging over time, we found that WT monolayers contain 79% more defects than Δ*pilH* monolayers. Fluctuations in the number of comet and trefoils closely follow one another, as predicted by nematic theory which requires that the total topological charge of the system must remain fixed[10]. **b**, This relationship was quantified further by calculating the normalized cross-correlation between comet and trefoil density. The maximum cross-correlation occurs at a lag time of zero for both strains, indicating that comet/trefoil pairs are created and annihilated instantaneously. This matches predictions made by previous SPR simulations[50]. **c**, Timeseries of Δ*pilH* and WT monolayers were processed to obtain measurements of the average flow of cells around comets and trefoils as for Fig. 3d. While the same characteristic flow structures were observed in both strains, we observed that the magnitude of the flow velocity was larger for the Δ*pilH* monolayer. This is consistent with Δ*pilH* monolayers having a larger activity than WT monolayers[11].

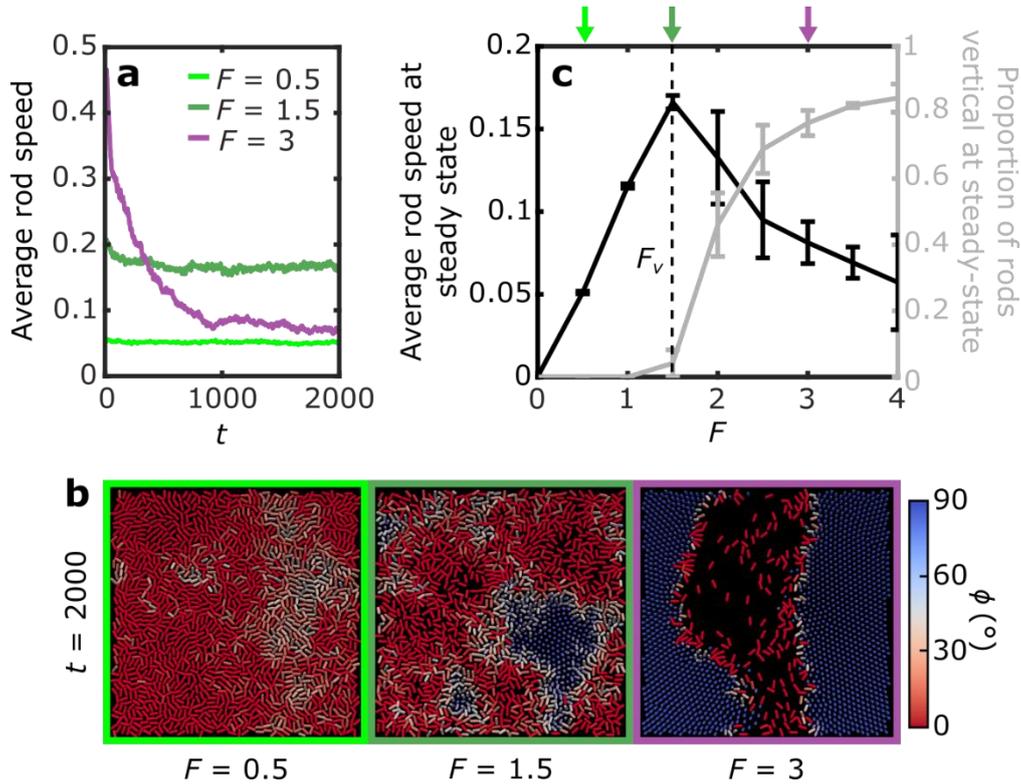

**Extended Data Fig. 9 | Rods that propel themselves with larger forces are more likely to become vertically oriented, which disrupts collective movement. a**, Measurements of the average rod speed as a function of time in three different simulations, each of which contains a uniform population of $N = 1600$ rods with an aspect ratio of $a = 4$ and a self-propulsive force, $F$. Although rods with $F = 1.5$ collectively move faster than rods with $F = 0.5$, increasing the propulsive force to $F = 3$ causes collective speed to sharply decline over time. **b**, Snapshots of simulations shown in **a** at steady state. Rods are color-coded by their orientation with respect to the surface, $\phi$, such that rods lying flat against the surface are shown in red, while those orthogonal to the surface are shown in blue. Rods with larger $F$ are more likely to stand on end, disrupting their capacity to move. **c**, We then performed independent simulations for different values of $F$ and plotted the average rod speed and proportion of rods oriented orthogonal to the surface at steady-state (Methods). This shows that the mean speed of the collective peaks at intermediate $F$, with larger values of $F$ causing rods to become vertically oriented. We denote the force that generates the maximum mean rod speed as $F_v$. Values of $F$ for simulations shown in **a** and **b** are denoted by coloured arrows. Our results show that out of plane cell rotation places an upper limit on how much propulsive force can be exerted within collectives. Lines and error bars show the mean and standard deviation of three simulations with different (random) initial conditions.

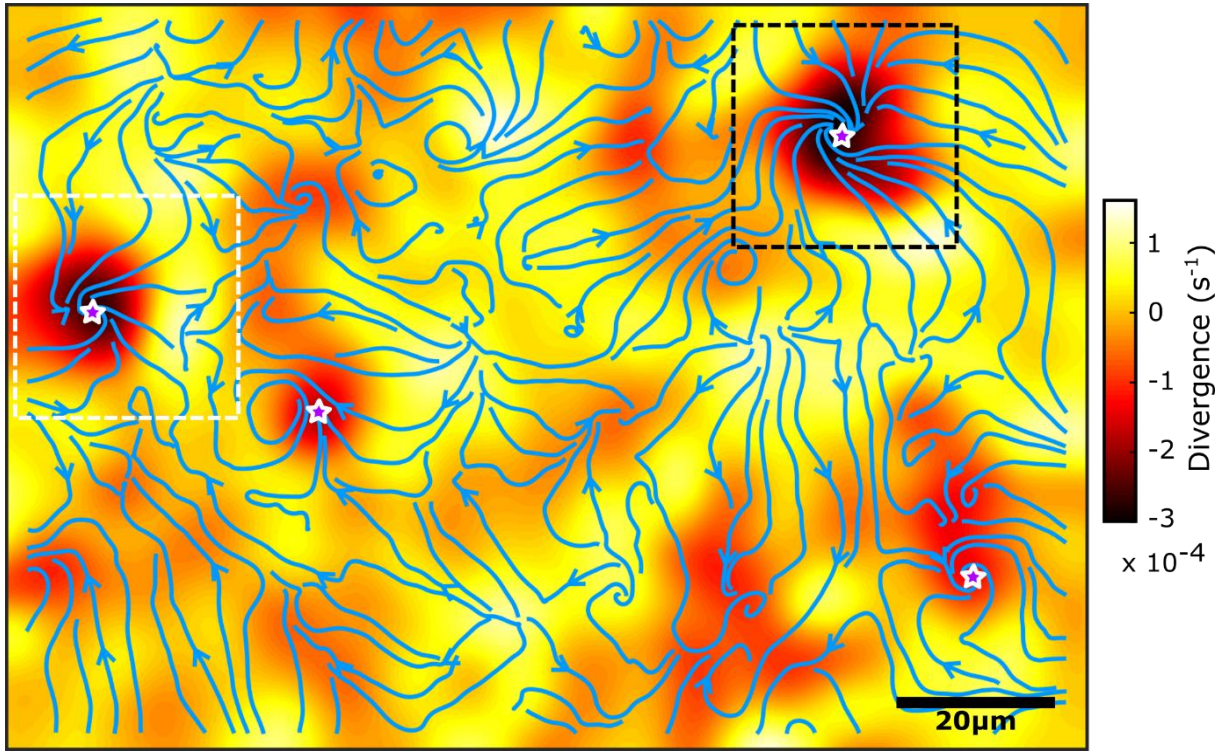

**Extended Data Fig. 10 | Rosettes drive points of convergent flow in a mixed WT/ΔpilH subsurface colony.** We used particle image velocimetry (PIV) to quantify the movement of cells across the two-dimensional coverslip over a period of 60 mins. This plot shows the streamlines and divergence of the temporally averaged velocity field. Stars denote locations where rosettes were observed to form. Regions where the divergence is negative indicate positions where cells locally accumulate. The region used for generating the first part of Supplementary Movie 9 and Fig. 4d-f is indicated with a black dashed box, the region used for the second part of Supplementary Movie 9 with a white dashed box.

# Supplementary Notes

Our experimental observations and modelling suggest that the increased force generated by Δ*pilH* cells causes them to become preferentially trapped in rosettes, which impairs their ability to spread (Fig. 4, Extended Data Figs. 9-10, Supplementary Movies 6-9). However, to rule out other possible mechanisms we performed a number of additional analyses. Specifically, we tested the following three alternative hypotheses:

1. Δ*pilH* cells grow more slowly than WT cells.
2. Δ*pilH* cells adhere to one another more strongly than WT cells.
3. The larger aspect ratio of Δ*pilH* cells disrupts their ability to spread.

Below, we discuss each of these hypotheses in turn.

## 1: Growth rate

Individual Δ*pilH* cells move more quickly than individual WT cells (Fig. 1g, Extended Data Fig. 3), likely because they express a larger number of pili on their surface[20,65]. Δ*pilH* cells exhibit increased levels of 3',5'-cyclic adenosine monophosphate (cAMP)[66], which is a secondary messenger known to increase pilus production, in addition to affecting a large number of other cellular processes[66,67]. Thus, it is plausible that Δ*pilH* cells could grow more slowly than WT cells. To investigate this possibility, we competed YFP-labelled Δ*pilH*, Δ*pilB* and WT cells with a CFP-labelled WT reference strain in liquid culture. While Δ*pilH* cells are hyperpilated compared to WT cells, Δ*pilB* cells lack pili on their surface[20]. In addition, we used the competition between the two differently labelled WT strains as a negative control. We counted the number of CFUs after 0, 3.5, and 7 h of incubation (Methods), allowing us to resolve the fitness of the YFP-labelled test strains relative to the WT reference strain over the same timescales as used in our subsurface colony assays (Extended Data Fig. 5a). Our results indicate that deletion of either *pilH* or *pilB* has no detectable impact on growth rate ($p > 0.05$, one-sample t-test, $n = 3$, Methods).

While Δ*pilH* cells did not have a growth rate defect in liquid culture, bacterial cells are exposed to different environmental conditions in colonies, which can affect gene expression[67,68]. In principle, this could mean that Δ*pilH* cells have a growth defect that manifests itself only in surface-based environments. However, the change to the intrinsic growth rate caused by the deletion of *pilH* is difficult to measure by scraping and enumerating whole colonies, as growth rate is confounded by the differential ability of cells to expand into new territory and acquire new resources (Fig. 2a, b). In order to show that differential growth alone cannot explain the slower expansion of Δ*pilH* cells, we estimated the growth rates required to produce the experimentally observed changes in the frequency of Δ*pilH* cells (Fig. 2f) assuming the two strains had an identical capacity to migrate.

We used a model of exponential growth to interpret our experimental data, using the same formalism used to describe population dynamics in chemostats[69]. We analysed the front and homeland as separate control volumes, assuming Δ*pilH* and WT cells grow at a constant rate given by $\mu_H$ and $\mu_W$, respectively. To see if only the difference in growth rate can explain the difference in expansion, we furthermore assumed that the migration of both strains was equivalent so they exit the control volume at a rate, $D$, that is equal for both strains. We can then model $X_H(t)$ and $X_W(t)$, the number of Δ*pilH* and WT cells respectively within the control volume, as:

$$\frac{dX_W(t)}{dt} = X_W(\mu_W - D), \tag{13a}$$

$$\frac{dX_H(t)}{dt} = X_H(\mu_H - D). \tag{13b}$$

Integrating with respect to time, $t$, yields:

$$X_W(t) = X_W(t_0)e^{(t-t_0)(\mu_W - D)}, \tag{14a}$$

$$X_H(t) = X_H(t_0)e^{(t-t_0)(\mu_H - D)}, \tag{14b}$$

where $t_0$ is some reference time. We can now write the fraction of $\Delta pilH$ cells, $f_H(t)$, as the size of the $\Delta pilH$ population divided by the total population size:

$$\begin{aligned} f_H(t) &= \frac{X_H(t_0)\, e^{\mu_H(t-t_0)} e^{-D(t-t_0)}}{X_H(t_0)\, e^{\mu_H(t-t_0)} e^{-D(t-t_0)} + X_W(t_0)\, e^{\mu_W(t-t_0)} e^{-D(t-t_0)}} \\ &= \frac{X_H(t_0)\, e^{\mu_H(t-t_0)}}{X_H(t_0)\, e^{\mu_H(t-t_0)} + X_W(t_0)\, e^{\mu_W(t-t_0)}}. \end{aligned} \tag{15}$$

Note that the $f_H(t)$ does not depend on $D$. We plot $f_H(t)$ in Fig. 2f for both the homeland and front regions of mixed WT/$\Delta pilH$ subsurface colonies.

Equation 15 can be rearranged to give the growth rate of the $\Delta pilH$ mutant in terms of the growth rate of the WT:

$$\mu_H = \frac{1}{(t-t_0)} \ln\left(\frac{f_H(t)\big(1 - f_H(t_0)\big)}{f_H(t_0)\big(1 - f_H(t)\big)} 2^{(t-t_0)/t_w}\right), \tag{16}$$

where $t_w = \frac{\ln(2)}{\mu_W}$ is the doubling time of WT cells. To fit this model to the "front" dataset in Fig. 2f, we considered the period from $t_0 = 200$ min to $t = 400$ min where $f_H(t)$ fell from 0.88 to 0.11 (Fig. 2f). For the purposes of this calculation, we assumed $t_w = 50$ min ($\mu_W = 0.0139$ min$^{-1}$), which was the doubling time of WT cells growing in liquid culture at the same temperature (Extended Data Fig. 5a). Substituting these values into (16) yields $\mu_H = -0.0072$ min$^{-1}$. This negative growth rate suggests the sharp decline in the fraction of $\Delta pilH$ at the front (Fig. 2f) cannot be explained purely by a difference in growth rate. We note that growth of WT cells in subsurface colonies is likely to be even slower than that observed in our liquid cultures, owing to nutrient competition between the densely packed cells. However, any increase in $t_w$ (the doubling time of the WT) will result in an even more negative value of $\mu_H$, which makes our calculation conservative. In conclusion, this analysis suggests that assuming an identical migration rate for the two strains is not correct, which is supported by observations that $\Delta pilH$ cells become preferentially trapped in rosettes at the same time their representation at the front decreases (Supplementary Movie 8).

In contrast to the front, $f_H$ gradually decreases in the homeland region. Considering that the decrease from $f_H = 0.62$ at $t_0 = 0$ to $f_H = 0.46$ at $t = 200$ min (Fig. 2f) and using the same WT growth rate estimate as above ($\mu_W = 0.0139$ min$^{-1}$, $t_w = 50$ min), we estimate the growth rate of $\Delta pilH$ cells to be $\mu_H = 0.0108$ min$^{-1}$ (equivalent to a doubling time of $t_H = 64$ min). This suggests that in contrast to the front, the decrease in the fraction of $\Delta pilH$ cells in the homeland could potentially be explained by a difference in intrinsic growth rate. However, this analysis likely underestimates growth rate of $\Delta pilH$ cells for two reasons. Firstly, we observe that the increased motility of $\Delta pilH$ cells causes them to preferentially migrate out of the homeland and accumulate in the leading edge of the colonies (Fig. 2f, g). This differential motility reduces the number of Δ*pilH* cells in the homeland during this period, a process which is not taken into account in our growth rate analysis. Secondly, Δ*pilH* cells' increased propensity to form rosettes results in cells being imaged from the end-on rather than in profile. Vertically oriented cells appear as small circles in our images, in comparison to horizontally oriented cells where the entire cell length is observed. As we estimate the fraction of each cell type using pixel-based measurements (Methods), rosette formation by Δ*pilH* cells in the homeland will act to reduce our estimate of $f_H$. We note that even without correcting for these two effects the measured growth rate difference between WT and Δ*pilH* cells in the homeland is insufficient to explain the rapid collapse of the Δ*pilH* population within the front. Taken together, these analyses suggest that even if Δ*pilH* does have a slight growth rate defect in subsurface colonies, it cannot explain the extent of the collapse of the Δ*pilH* population in the front at $t = 300$ min.

**2: Cell-cell adhesion**

Our experiments show that motility driven by Type IV pili greatly enhances the expansion rate of *P. aeruginosa* colonies (Fig. 1e, f, Supplementary Movie 3). However, Type IV pili have also been shown to increase cell-cell adhesion in both colonies of *Neisseria gonorrhoeae*, which are composed of spherical cells that expand primarily by cell division, and in rafts of swarming *P. aeruginosa*, which use flagella to collectively swim in a thin film of liquid[23,24]. We therefore tested if intercellular adhesion played a role in the accumulation of Δ*pilH* cells within rosettes. Because Δ*pilH* cells are hyperpilated, one might expect them to adhere to one another more strongly than WT cells, preferentially inhibiting their outwards migration. To investigate the role of cell-cell adhesion in our colony assays, we studied the movement of Δ*pilH* cells at low density so that we could resolve the interaction between individual cells.

We tracked the movement of cells three hours after they were inoculated into a subsurface colony. At this stage, cells were highly motile but the majority were spaced far apart from one another. We then isolated (*i*) events where previously separated cells come into contact with one another and (*ii*) events where cells already in contact move away from one another. If cells actively adhered to each other, we would expect that cells would both slow down after contacting one another and increase their speed after moving away from one another.

We manually isolated 41 events where a previously isolated cell came into contact with another cell and 47 events where two cells already in contact moved apart (Extended Data Fig. 5b). We then measured the speed of cells undergoing these events. The timeseries were aligned with each another so that the timepoint at which contact was made or broken occurred at $t = 0$, and then averaged the speed of cells undergoing each type of event.

For both categories of cell movement, we observed a peak in cell speed at $t = 0$ because movement is required for either event to be detected (Extended Data Fig. 5b). Importantly however, there was no discernible difference in the speed of cells before and after either event. Moreover, the time series from cells making contact and breaking contact with one another are indistinguishable, suggesting that cell adhesion is negligible. Visual inspection of individual cells also failed to reveal any evidence of aggregation or changes in motility behaviour during cell-cell contact. These results suggest that cell-cell adhesion does not play a substantial role in the collective movement of Δ*pilH* cells.

This finding is consistent with our observation that Δ*pilH* cells in density packed collectives move faster than solitary ones (Extended Data Fig. 3), indicating that the hyperpilation of Δ*pilH* cells enhances collective motility rather than stifling it through enhanced cell-cell adhesion. Moreover, our experiments indicate defect collision drives an inward migration of cells towards points that ultimately form rosettes (Fig. 4d-f, Supplementary Movie 9), suggesting that rosette formation involves active motility rather than passive aggregation driven by increased adhesion.

### 3: Aspect ratio

We observed that Δ*pilH* cells appeared to be somewhat longer than WT cells when each were in monoculture (e.g. Supplementary Movie 5). To verify this, we imaged both genotypes when mixed together in a single monolayer, ensuring that both experienced the same environmental conditions. We used scanning laser confocal fluorescence microscopy to identify the differently labelled genotypes and measured the size of cells by fitting them with ellipses. This analysis confirmed our initial observation: Δ*pilH* cells had an average length of 3.64 μm ($n = 217$), whereas WT cells had an average length of 3.12 μm ($n = 223$). The difference in measured cell length was significant ($p < 10^{-10}$, Mann-Whitney *U* test). Similar findings were observed for (*i*) solitary cells at low density in the subsurface assay and (*ii*) in liquid cultures growing at exponential phase (Extended Data Fig. 6a). Taken together, these analyses show that Δ*pilH* cells are longer than WT cells across a wide range of conditions. While the mechanisms involved are not known, we speculate the enhanced cAMP concentrations found in Δ*pilH* cells[66] may play a role.

Cell length has been observed to impact collective behaviours that arise from both passive growth[70] and active motility[71] in colonies of rod shaped bacteria. To understand how the differences in cell length impact rosette formation we used our 3D SPR model to simulate monolayers composed of cells with different aspect ratios ($a = 3.5, 4$, and $5$), spanning the range observed in our experiments (Extended Data Fig. 6a). We varied the force, $F$, exerted by each of these different rods and quantified both the mean rod speed and fraction of vertical rods at steady state (Extended Data Fig. 6b, c). These results show that longer rods have a *decreased* tendency to reorient out of plane for a given $F$ (Extended Data Fig. 6c). Consequently, monolayers composed of longer rods can propel themselves with a larger force, $F$, before they begin to buckle out of plane, allowing them to achieve a larger mean speed (Extended Data Fig. 6b). Mechanistically, this occurs because rotating a longer cell out of plane requires a larger deformation of the elastic materials that act to stabilize horizontal orientation of the cells (i.e. the overlaying agar and the polymeric secretions that glue cells to surfaces). We predict that the torque required to rotate a cell into a vertical orientation increases with its length squared (equation 8), suggesting that small increases in cell length could dramatically stifle rosette formation.

Our analyses therefore indicate that the increased length of Δ*pilH* cells actually make it harder for them to rotate out of plane and form rosettes. Thus, the increased force generated by Δ*pilH* cell causes them to preferentially form rosettes *in spite* of their longer length. This finding is consistent with previous work on non-motile *Vibrio cholerae* colonies, which also found that longer cells are more resistant to verticalization than shorter ones[27].

To confirm that these findings from monoculture simulations could be extended to mixed genotype simulations, we performed additional SPR simulations that model the interaction between cells with different lengths. To approximate WT and Δ*pilH* cells, we considered rods with an aspect ratio of $a_1 = 4$ and $a_2 = 5$, which approximate experimental measurements of the WT and Δ*pilH*, respectively (Extended Data Fig. 6a). In the simulations presented in Fig. 4b, we quantified the interactions between rods with the same aspect ratio and different $F$. In the simulations presented in Extended Data Fig. 6d, e, we kept the force generated by the "WT cells" constant and varied the force generated by the longer "Δ*pilH* cells" (i.e. "WT cells" have fixed aspect ratio $a_1 = 4$ and self-generated force $F_1 = 1.5$, whereas "Δ*pilH* cells" have fixed $a_2 = 5$ and exert a variable force, $F_2$). As anticipated from the monoculture simulations, in cases that $F_2$ exceeded $F_1$ by a small amount the longer rods were able to move faster than the shorter rods on average. However, once $F_2$ increased beyond a critical threshold the longer cells were preferentially trapped within rosettes, which reduced their relative mean speed. Compared to our simulations where both populations had the same length ($a_1 = a_2 = 4$, Fig. 4b), we observed that rosette formation was triggered at a larger value of $F_2$ owing to the increased stability of the longer "Δ*pilH* cells" ($a_1 = 4, a_2 = 5$, Extended Data Fig. 6d). Interestingly, we found that once the critical value of $F_2$ is exceeded, larger aspect ratio actually *increases* the representation of "Δ*pilH* cells" in rosettes (Extended Data Fig. 6d, e). In other words, a larger aspect ratio allows cells to exert more force before forming rosettes, but once this critical threshold is crossed the longer cells are more likely to become trapped in the resulting rosettes than the shorter ones. This observation is consistent with our experiments, which show that the higher force and longer Δ*pilH* cells are much more likely to end up in rosettes than the lower force and shorter WT cells (Fig. 4f). The increased representation of longer cells within rosettes is also consistent with the longer cells having a stronger tendency to align with their neighbours, which in this case are vertically oriented. Similar effects are widely observed in the context of nematic alignment in liquid crystals, in which a larger aspect ratio leads to a larger orientational elasticity[72,73] and therefore stronger alignment.

In conclusion, we observe that Δ*pilH* cells are longer than WT cells. The greater length of Δ*pilH* cells stabilizes them against rosette formation, so they have to generate an even larger force before they can trigger rosettes. However, once rosettes form, the longer length of Δ*pilH* cells increases their representation in rosettes relative to the shorter WT cells.

Taken together, our supplementary analyses strongly support the conclusion that the higher force generated by Δ*pilH* cells is the primary mechanism responsible for the formation of rosettes.

# Supplementary Movies

**Supplementary Movie 1 | *P. aeruginosa* displays collective motility in surficial colonies**. A surficial colony was inoculated with equal proportions of YFP and CFP labelled WT cells. Shown here is the monolayer after 48 h of incubation at room temperature. The left image shows both strains imaged using brightfield microscopy. The right panel shows only the YFP labelled cells. Patterns of collective movement are clearly visible in the right-hand image because only half of the cells are visible. Total duration is 20 mins.

**Supplementary Movie 2 | Our custom tracking software (FAST) can follow the movement of individual cells even when very tightly packed together.** Cyan dots show cell centroids and orange traces show cell trajectories. Total duration is 2 min.

**Supplementary Movie 3 | Subsurface Δ*pilH* colonies initially expand more quickly than WT colonies, but are eventually overtaken.** Brightfield images were background subtracted, inverted and contrast enhanced (Methods). Four adjacent fields of view were recorded and stitched together to form a single image. The dark vertical lines are caused by subtle variations in focus between adjacent fields of view. A colony of Δ*pilB* cells, which lack pili-based motility, is shown as a control. Total duration is 15 h.

**Supplementary Movie 4 | Quantification of bacterial competition in subsurface colonies using automated image analysis.** WT and Δ*pilH* cells were mixed together in equal fractions and used to inoculate a subsurface colony. Automated routines were then used to resolve the location of both the "front" (green boxes) and "homeland" (purple boxes), which were then used in subsequent analyses (Fig. 2d-f). While these datasets also include fluorescent images (Fig. 2g), shown here is the brightfield channel, which is processed as described in the Methods section. Total duration is 8 h.

**Supplementary Movie 5 | Automated detection of topological defects in WT and Δ*pilH* monolayers.** Red circles indicate locations of comet defects (+1/2 charge) and blue triangles trefoil defects (-1/2 charge). Orange arrows and cyan lines indicate the orientation of comets and trefoils, respectively. Total duration is 5 min.

**Supplementary Movie 6 | The collision of comets generates stable rosettes only if the force exerted by self-propelled rods exceeds a critical threshold.** We initialized our 3D SPR model with two comets (red circles, orange arrows) directed towards one another. The initial static image shows the initial configuration of rods. The trefoils (blue triangles) ensure the total topological charge of the system is zero. Rods that propel themselves with a relatively small force ($F = 1$, left) remain horizontally oriented after an initial transient, whereas rods that propel themselves with a larger force ($F = 3$, right) form stable rosettes. Snapshots from these simulations are presented in Fig. 4a.

**Supplementary Movie 7 | Increased propulsive force causes rosettes to form in a monolayer of rods initialized in a random configuration.** Three separate 3D SPR simulations are shown. Rods propel themselves with $F = 0.5$, $F = 1.5$, and $F = 3$ in the left, middle, and right simulations respectively. All other parameters were kept constant between simulations. Data from these simulations are shown in Extended Data Fig. 9.

**Supplementary Movie 8 | Rosettes in mixed WT and Δ*pilH* colonies are primarily composed of Δ*pilH* cells**. The left panel shows the monolayer of a subsurface colony inoculated with an equal proportion of CFP-labelled WT cells (grey) and YFP-labelled Δ*pilH* cells (yellow). The right image shows an analogous experiment, but with the opposite labelling i.e. with Δ*pilH* cells (in grey) and WT cells (in yellow). The first half of this movie shows an overlay of bright field and YFP channels (showing both cell types), while the latter half shows only the YFP-labelled cells. Δ*pilH* cells in both experiments become preferentially trapped in rosettes, which appear as yellow patches and dark patches in the left and right images respectively. To minimize phototoxicity, we did not image the CFP channel in these experiments. Total duration is 3 h.

**Supplementary Movie 9 | The collision of comets triggers rosette formation in a monolayer composed of both WT and Δ*pilH* cells.** This movie shows the formation of two different rosettes, in turn. The initial static image in each sequence shows the position of comets (red circles, orange arrows) and trefoils (blue triangles, cyan spokes) prior to rosette formation. In both cases, rosettes are initiated by the collision of two comets. The flowfield generated during the formation of both rosettes is shown in Extended Data Fig. 10. The first half of the movie shows the formation of the rosette presented in Fig. 4d-f; subsequent confocal imaging revealed this rosette was nearly wholly composed of vertically oriented Δ*pilH* cells. Duration of each sequence is 60 min.